\documentclass[twocolumn,prd,nofootinbib,lengthcheck,superscriptaddress,letterpaper,showpacs]{revtex4-1} %
\usepackage{hyperref}
\usepackage{verbatim}
\usepackage{graphicx}
\usepackage{dcolumn}
\usepackage{bm}
\usepackage[usenames,dvipsnames]{color}
\usepackage{url}
\usepackage{amsmath}
\usepackage{times}
\usepackage[varg]{txfonts}
\usepackage{longtable}

\newcommand\qmstateproduct[2]{\left\langle#1|#2\right\rangle}

\newcommand\Y[1]{{{}_{#1}Y}}

\newcommand\lnLmarg{ \ln {\cal L}_{\rm marg}}
\newcommand\lnLmargMax{ 272.5}
\newcommand\lnLmargGray{265.8 }
\newcommand\lnLmargBlack{ 268.6}
\newcommand\pCI{90\%}
\newcommand\ForInternalReference[1]{}
\newcommand\TheColorSEOB{black}
\newcommand\TheColorIMRP{blue}

\newcommand{\macro}[1]{\textcolor{red}{#1}} 

\newcommand\OBSEVENTFULLDATE{\macro{September 14, 2015 09:50:45 UTC}}

\newcommand{\MTOTobsRANGE}{\macro{\ensuremath{66\text{--}75}}} 
\newcommand{\MONEobsRANGE}{\macro{\ensuremath{35\text{--}45}}} 
\newcommand{\MTWOobsRANGE}{\macro{\ensuremath{27\text{--}36}}} 
\newcommand{\MFINALobsRANGE}{\macro{\ensuremath{63\text{--}71}}} 
\newcommand{\MASSRATIORANGE}{\macro{\ensuremath{0.62\text{--}0.98}}} 
\newcommand{\CHIEFFRANGE}{\macro{\ensuremath{-0.24\text{--}0.09}}} 
\newcommand{\SPINONERANGE}{\macro{\ensuremath{0.0\text{--}0.8}}} 
\newcommand{\SPINTWORANGE}{\macro{\ensuremath{0.0\text{--}0.9}}} 
\newcommand{\SPINFINALRANGE}{\macro{\ensuremath{0.60\text{--}0.72}}} 
\newcommand{\PEOPTIMALSNRRANGE}{\macro{\ensuremath{23.5\text{--}26.8}}} 

\newcommand{\CONSTRAINTALIGNEDSPINONELIMIT}{\macro{\ensuremath{0.2}}} 
\newcommand{\CONSTRAINTALIGNEDSPINTWOLIMIT}{\macro{\ensuremath{0.3}}} 
\renewcommand{\CONSTRAINTALIGNEDSPINONELIMIT}{\macro{\ensuremath{0.22}}} %
\renewcommand{\CONSTRAINTALIGNEDSPINTWOLIMIT}{\macro{\ensuremath{0.25}}} %
\renewcommand{\macro}[1]{#1}

\definecolor{amber}{rgb}{1.0, 0.75, 0.0}
\definecolor{orange}{rgb}{1.0, 0.5, 0.0}
\definecolor{amaranth}{rgb}{0.9, 0.17, 0.31}

\newcommand\unit[1]{\,{\rm #1}}
\usepackage[normalem]{ulem} %

\graphicspath{{./figures/}}
\newcommand{\mc}{{\cal M}}

\def\ltsima{$\; \buildrel < \over \sim \;$}
\def\simlt{\lower.5ex\hbox{\ltsima}}
\def\gtsima{$\; \buildrel > \over \sim \;$}
\def\simgt{\lower.5ex\hbox{\gtsima}}

\def\apj{Astrophysical Journal}                %
\def\apjl{Astrophysical Journal}             %

\def\prd{Phys.~Rev.~D}       %
\def\prl{Phys.~Rev.~Lett}    %
\def\araa{ARA\&A}             %
\def\aap{A\&A}                %

\def\TheEvent{GW150914}

\newcommand\ReportedNumberThisPaperMfzMsunLow{64.0}
\newcommand\ReportedNumberThisPaperMfzMsunHigh{73.5}
\newcommand\ReportedNumberThisPaperAFLow{0.62}
\newcommand\ReportedNumberThisPaperAFHigh{0.73}
\newcommand\SimulationsProvidedNumber{1139}
\newcommand\SimulationsProvidedNumberRIT{394}
\newcommand\SimulationsProvidedNumberSXS{310} %
\newcommand\SimulationsProvidedNumberGT{406}    %
\newcommand\SimulationsProvidedNumberBAM{29}
\newcommand\SimulationsProvidedNumberFollowupTotal{108}  %
\newcommand\SimulationsProvidedNumberFollowupRIT{52}
\newcommand\SimulationsProvidedNumberFollowupSXS{8}

\newcommand\PaperDetection{\underline{LVC-detect}\cite{GW150914-DETECTION}}  %
\newcommand\PaperPE{\underline{LVC-PE}\cite{GW150914-PARAMESTIM}}%
\newcommand\PaperTestGR{\underline{LVC-TestGR}\cite{GW150914-TESTOFGR}}%
\newcommand\PaperPENRMethods{\underline{PE+NR-Methods}\cite{gwastro-PENR-Methods-Lange}}
\newcommand\PaperAstro{\underline{LVC-Astro}\cite{GW150914-ASTRO}}%
\newcommand\PaperBurst{\underline{LVC-Burst}\cite{GW150914-BURST}}%

\newcommand\PaperSEOBNRvthree{\underline{LVC-SEOBNRv3}\cite{SEOBv3Paper}}

\begin{document}

\title{Directly comparing \TheEvent{} with numerical solutions of Einstein's equations for  binary black hole coalescence}
\author{B.\,P.~Abbott \emph{et al.}}
\email{Full author list given at the end of the article}%
\noaffiliation{}

\today

\begin{abstract}
We compare  \TheEvent{} directly to simulations of coalescing binary
black holes in full general relativity, including   several performed specifically to reproduce this event.  
Our calculations go beyond existing semianalytic models, because for all simulations --  including sources with two independent, precessing
spins -- we perform comparisons which account for all the spin-weighted quadrupolar  modes, and separately which account
for all the quadrupolar and octopolar  modes.
Consistent with the posterior distributions reported in \PaperPE{} (at the \pCI{} credible level), we find the data are compatible with a wide range of
nonprecessing \emph{and  precessing} simulations.   
Followup simulations performed using  previously-estimated binary parameters most resemble the data,
even when all quadrupolar and octopolar modes are included.  
Comparisons  including only the quadrupolar modes    constrain the total redshifted mass $M_z \in [64 M_\odot-82M_\odot]$, mass ratio $1/q=m_2/m_1 \in[0.6,1]$,
and effective aligned spin $\chi_{\rm eff} \in [-0.3, 0.2]$, where $\chi_{\rm eff} = (\mathbf{S}_1/m_1
  +\mathbf{S}_2/m_2)\cdot \mathbf{\hat{L}}/M$.    Including both quadrupolar and  octopolar modes, we find the 
mass ratio is even more tightly constrained.
 Even accounting for precession, simulations with extreme mass ratios and
effective spins are highly inconsistent with the data, at any mass.  
Several nonprecessing and precessing simulations with similar mass ratio and $\chi_{\rm eff}$ are consistent with the
data.
Though correlated, the components' spins (both in  magnitude and directions) are not significantly constrained
  by the data:  the data is consistent with simulations with component spin magnitudes
   $a_{1,2}$ up to at least $0.8$, with random orientations.
Further detailed followup calculations are needed to determine if  the data  contain a weak imprint from transverse (precessing) spins.
For nonprecessing binaries,  interpolating between simulations, we reconstruct a posterior distribution  consistent with
previous results. 
The final black hole's redshifted mass is consistent with $M_{f,z}  $ in the range
$\ReportedNumberThisPaperMfzMsunLow M_\odot-\ReportedNumberThisPaperMfzMsunHigh M_\odot$ and the final black hole's dimensionless spin parameter is consistent with $a_f=\ReportedNumberThisPaperAFLow-\ReportedNumberThisPaperAFHigh$. %
As our approach invokes no intermediate approximations to general relativity and can strongly reject binaries whose
radiation is inconsistent with the data, our
analysis provides a valuable complement to \PaperPE{}.
\end{abstract}
\maketitle

\section{Introduction}

On \OBSEVENTFULLDATE{}, gravitational waves were observed in coincidence
by the twin instruments of the Laser Interferometer Gravitational-wave
Observatory (LIGO) located at Hanford, Washington, and Livingston, Louisiana, in the USA, an event known as
\TheEvent{}\cite{GW150914-DETECTION}.  %
\PaperDetection{}, \PaperPE{}, \PaperTestGR{}, and  \PaperBurst{} demonstrate  consistency between \TheEvent{} and
selected individual predictions for a binary
black hole coalescence, derived using 
numerical solutions of Einstein's equations for general relativity. %
\PaperPE{} described a systematic, Bayesian method to reconstruct the properties of the coalescing binary, by comparing the data  with the expected
gravitational wave signature from  binary black hole coalescence \cite{gw-astro-PE-lalinference-v1},  evaluated using state-of-the-art semianalytic approximations 
to its dynamics and
radiation \cite{ 2014PhRvD..89f1502T,2014CQGra..31s5010P,gwastro-mergers-IMRPhenomP}.    %

In this paper, we present an alternative method of reconstructing the 
binary parameters
of \TheEvent{}, without using the semianalytic waveform
models employed in \PaperPE{}.  Instead, we compare
the data directly
with the most physically complete and generic
predictions of general relativity: computer simulations of binary black hole coalescence in full nonlinear general
relativity (henceforth referred to as numerical relativity, or NR).   
Although the semianalytic models are calibrated
to NR simulations, even the best available models only imperfectly reproduce the predictions of numerical relativity, on a
mode-by-mode basis \cite{gwastro-mergers-nr-PrayushAlignedSpinAccuracy-2016}.
Furthermore, typical implementations of these models, such as
those used in \PaperPE{}, consider only the
dominant %
spherical-harmonic mode of the waveform (in a corotating frame).
For all NR simulations considered here---including sources with two independent, precessing
spins---we perform comparisons that account for all the quadrupolar  %
spherical-harmonic waveform modes, and separately comparisons that account for all the quadrupolar and octopolar spherical harmonic modes.  

The principal approach introduced  in this paper is  different from \PaperPE{}, which  inferred the properties
of \TheEvent{} by adopting analytic waveform models.   Qualitatively speaking, these models  interpolate the \emph{outgoing gravitational wave
  strain} (waveforms)  between the well-characterized results
of numerical relativity, as provided by a sparse grid of simulations.     These interpolated or analytic waveforms are used to
generate a continuous  posterior distribution over the   binary's parameters.   By contrast, in this study, we 
compare numerical relativity to the data first, evaluating  a single scalar quantity (the marginalized likelihood)  on
the grid of binary parameters prescribed and provided by all available NR simulations.  We then construct 
  an approximation to the marginalized likelihood that interpolates between NR simulations with different parameters.
 To the extent that the likelihood is a  simpler function of parameters than the waveforms, this method
  may require fewer NR simulations and fewer modeling
  assumptions. Moreover, the interpolant for the likelihood needs to be
  accurate only near its peak value, and not everywhere in parameter space.
A similar study was conducted on GW150914 using a subset of numerical relativity waveforms directly against
reconstructed waveforms~\cite{GW150914-BURST}; the results reported here are consistent but more thorough.

Despite using an analysis that has few 
features or code in common with the methods employed in \PaperPE{},
we arrive 
at similar conclusions regarding the
parameters of the progenitor black holes and the final remnant,
although we extract slightly more information about 
the binary mass ratio by using higher-order modes.
Thus, we provide independent corroboration of the
results of \PaperPE{}, strengthening our confidence in both the
employed statistical methods and waveform models.

This paper is organized as follows.  Section \ref{sec:Motivation} provides an extended motivation for and 
summary of this
investigation.   Section \ref{sec:sub:NRHistory} reviews the history of numerical relativity and introduces notation to
characterize simulated binaries and their radiation.  Section \ref{sec:Methods:NR:Sims} describes the simulations used in this work. 
Section \ref{sec:Methods:PE} briefly describes the method used to compare simulations to the data; see
\PaperPENRMethods{} for further details.  Section \ref{sec:NR:DurationIssues} describes the implications of using NR simulations
that include only a small number of gravitational-wave
cycles.   Section \ref{sec:Compare} relates this investigation to prior work.  
Section  \ref{sec:Results} describes our results on the pre-coalescence parameters.  We provide a ranking of simulations as measured by a simple measure of
fit (peak marginalized log likelihood).  When possible, we provide an approximate posterior distribution over all
intrinsic parameters.   Using both our simple ranking and approximate posterior distributions, we draw conclusions
about the range of source parameters that are consistent with the data.
Section \ref{sec:Results:FinalState} describes our results on the post-coalescence state.
Our statements rely on the
final black hole masses and spins derived from the full NR simulations 
used.
We summarize our results in Section \ref{sec:Conclude}.  
In Appendix \ref{ap:FullSimulationDetails}, we summarize the simulations used in this work and their accuracy, referring to the original
literature for complete details.

\section{Motivation for this study}
\label{sec:Motivation}

\begin{figure}
\includegraphics[angle=270,width=\columnwidth]{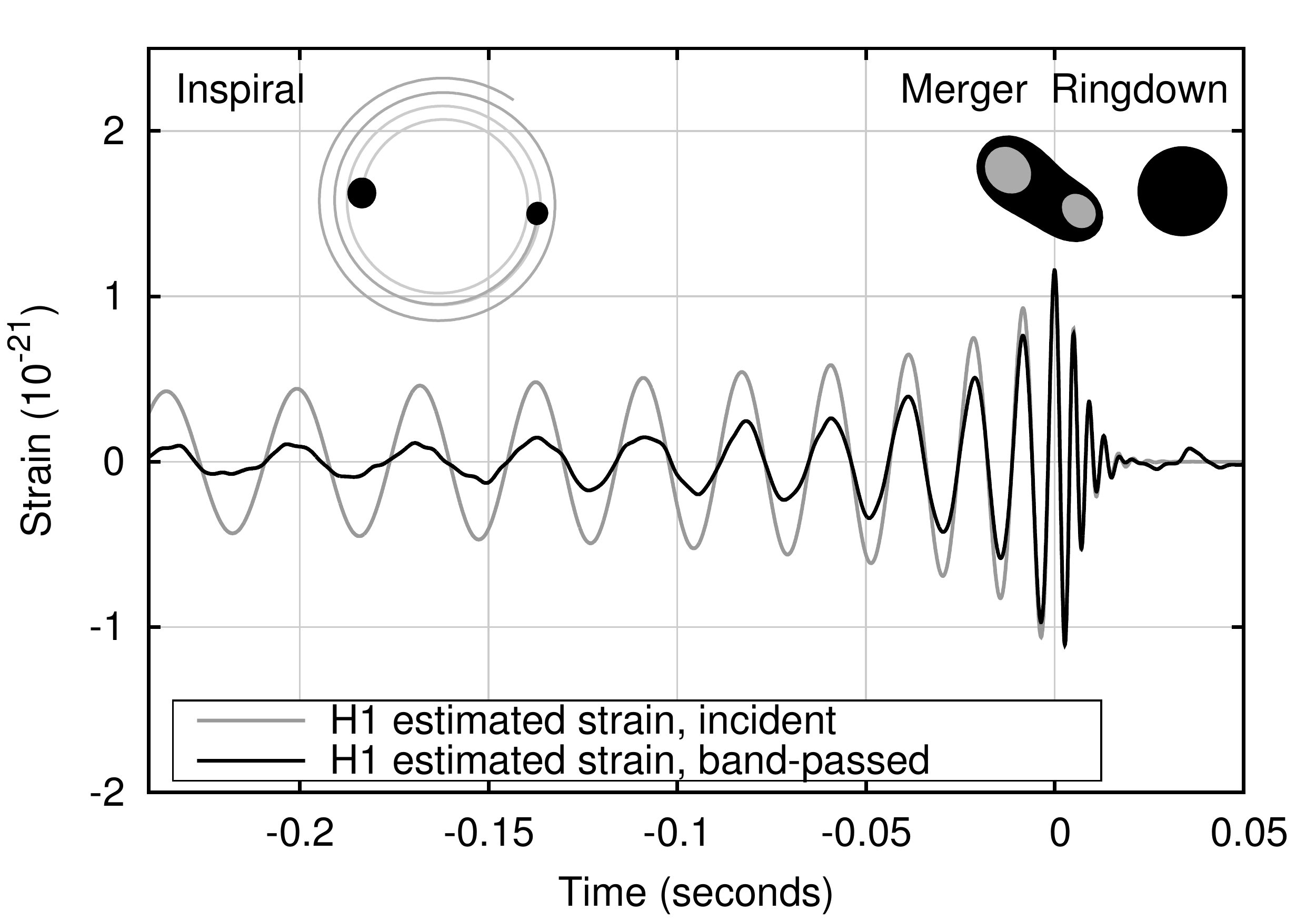}
\caption{\label{fig:Intro}\textbf{Simulated waveform}:  
Predicted strain in H1 for a source with parameters
  $q=1.22,\chi_{1,z}=0.33,\chi_{2,z}=0.44$, simulated in full general relativity; compare to Figure 2 in \PaperDetection{} .  The gray line shows the idealized strain response
  $h(t)=F_+ h_+(t)+F_\times h_\times(t)$, while the solid black line shows the whitened strain response, using the same
  noise power spectrum as \PaperDetection.
}
\end{figure}

This paper presents an alternative analysis of \TheEvent{} 
and an alternative determination of its intrinsic parameters.  
The methods
used here differ from those in \PaperPE{} in two important ways. First,
the statistical analysis here is performed in a manner different than and
independent of the one in \PaperPE{}.  Second, the gravitational waveform 
models used in \PaperPE{} are analytic approximations of particular
functional forms, with coefficients calibrated to match selected
NR simulations; in contrast, here we directly use waveforms
from NR simulations.
Despite these differences, 
our conclusions largely corroborate the quantitative results found in
\PaperPE{}.
Our study also addresses key challenges associated with 
gravitational wave parameter estimation for  black
hole binaries with total mass $M> 50 M_\odot$.  In this mass regime, LIGO is sensitive to the last few dozens of cycles
of coalescence, a strongly nonlinear epoch that is the most difficult to approximate with analytic  (or semi-analytic) waveform models 
\cite{2015PhRvD..92j2001K,2015PhRvL.115l1102B,2014PhRvD..89f1502T,LIGO-Puerrer-NR-LI-Systematics}.   
 Figure \ref{fig:Intro} illustrates the dynamics and expected detector response in this
regime, for a source like \TheEvent{}.  
For these last dozen cycles, existing analytic waveform models 
have only incomplete descriptions of precession,  lack higher-order
spherical-harmonic modes, and do not fully account for strong
nonlinearities.  
Preliminary investigations have shown that  inferences about the source drawn using these existing analytic approximations can  be slightly or significantly biased
\cite{gwastro-mergers-nr-PrayushAlignedSpinAccuracy-2016,LIGO-Puerrer-NR-LI-Systematics,gwastro-mergers-nr-Alignment-ROS-CorotatingWaveforms,2014PhRvD..90l4004V}.  
Systematic studies are underway to assess how these approximations influence our best estimates of a candidate
  binary's parameters.  At present, we can only summarize the rationale for these investigations, not their  results.  
To provide three concrete examples of omitted physics, first and foremost, even the most sophisticated models for binary black hole coalescence
\cite{2014PhRvD..89f1502T} do not yet account for the asymmetries \cite{gwastro-mergers-nr-Alignment-ROS-CorotatingWaveforms}  responsible for the largest gravitational-wave recoil
kicks \cite{2008PhRvD..77d4031S,2015PhRvD..92b4022Z}. %
Second,  the  analytic waveform models adopted in \PaperPE{}
 adopted simple spin treatments (e.g., a binary
with aligned spins, or a binary with single effective precessing spin)
that cannot capture the full spin
dynamics~\cite{ACST,gwastro-mergers-PNLock-Morphology-Kesden2014,gwastro-mergers-PNLock-PRLFollowup}. 
A single precessing spin is often a good approximation, particularly for unequal masses where one spin dominates the
angular momentum budget \cite{ACST,2012PhRvD..86f4020B,2014PhRvD..89d4021L,gwastro-mergers-IMRPhenomP,2015PhRvD..91b4043S}.  
However, for appropriate comparable-mass sources, two-spin effects are known to be  observationally accessible
\cite{gwastro-mergers-nr-Alignment-ROS-CorotatingWaveforms,gwastro-mergers-PNLock-Distinguish-Daniele2015} and cannot be
fully captured by a single spin. 
Finally, \PaperPE{} and \PaperTestGR{} made an additional approximation to connect the inferred properties of
the binary black hole with the final black hole state \cite{Healy:2014yta,2015arXiv150807250H}.
The analysis presented in this work does not rely on these
approximations: observational data are directly compared against a 
wide range of NR simulations
and the final black
hole properties are extracted directly from these simulations, 
without recourse to estimated relationships
between the initial and final state.  
By circumventing these approximations,
our analysis can corroborate conclusions about selected 
physical properties of \TheEvent{} presented in  those papers.  
Despite the apparent simplicity of \TheEvent{}, we find that
a range of binary black hole masses
and spins, including strongly precessing systems with significant misaligned
black hole spins~\cite{2015PhRvL.114n1101L}, are
reasonably consistent with the data.
The reason the data cannot distinguish between sources with
qualitatively different dynamics is a consequence of both
the orientation of the source relative to the line of sight and 
the timescale of \TheEvent{}.
First,
if the line of sight is along or opposite
the total angular momentum vector of
the source, even the most strongly precessing
black hole binary emits a weakly-modulated inspiral signal, 
lacking unambiguous signatures of precession and easily
mistaken for a nonprecessing binary~\cite{gwastro-mergers-PNLock-Distinguish-Daniele2015,gwastro-pe-Tyson-AstroSample-MassGap2015}.  Second, because \TheEvent{} has a large
total mass, very little of the inspiral lies in LIGO's frequency
band, so the signal is short, with few orbital cycles and even 
fewer precession cycles prior to or during
coalescence.   
The short duration of \TheEvent{} provides few opportunities for 
the dynamics of two precessing spins to introduce 
distinctive amplitude and phase modulations into
its gravitational wave inspiral signal 
\cite{ACST}.   

Although the orientation of the binary and the short duration of the signal
make it difficult to extract spin information from the {\em inspiral},
comparable-mass binaries with large spins can have exceptionally
rich dynamics
with waveform
signatures that extend into the late inspiral and the strong-field
merger phase 
\cite{gwastro-mergers-nr-Alignment-ROS-Polarization,gwastro-mergers-IMRPhenomP}.
By utilizing full NR waveforms instead of the single-spin (precessing) and 
double-spin (nonprecessing) models applied in \PaperPE{}, the approach
described here provides an independent opportunity to extract additional 
insight from the data, or to independently corroborate the
results of \PaperPE{}.
Our study employs a simple method to carry out our Bayesian calculations: for each NR simulation, we evaluate the marginalized
likelihood of the simulation parameters given the data.  The likelihood is evaluated via an adaptive Monte Carlo integrator.    This method provides a quantitative ranking of simulations; with judicious
interpolation in parameter space,  the method also allows us to identify candidate 
parameters for followup numerical relativity simulations.
To estimate parameters of \TheEvent{}, we can simply select
the subset of
 simulations and masses that have a marginalized likelihood greater than an observationally-motivated threshold (i.e., large enough to
 contribute significantly to the posterior).   
Even better, with a modest approximation to fill the gaps between NR simulations, we can reproduce and corroborate the
results in \PaperPE{} with a completely independent method.
 We explicitly construct an approximation to the likelihood
that interpolates between simulations of precessing binaries, and demonstrate
its validity and utility.
It is well-known that the choice of prior may influence conclusions
of Bayesian studies when the data do not strongly constrain the
relevant parameters. For example, the results of
\PaperPE{} suggest that \TheEvent{} had low to moderate spins,
but this is due partly to the conventional
prior used in \PaperPE{} and earlier studies
\cite{gw-astro-PE-lalinference-v1}. This prior 
is uniform in spin \emph{magnitude}, 
and therefore unfavorable to
the most dynamically interesting
possibilities: comparable-mass binaries with two large, 
dynamically-significant precessing spins
\cite{gwastro-mergers-PNLock-Distinguish-Daniele2015}.
In contrast, by directly 
considering the (marginalized) likelihood, the results of our study
are independent of the choice of prior.  For example, we find here
that \TheEvent{}
is consistent with two large, dynamically significant spins.
Finally, our efforts to identify even subtle hints of spin precession are motivated by the astrophysical opportunities afforded
by spin measurements with \TheEvent; see \PaperAstro.    Using the geometric spin prior adopted in \PaperPE, the data from 
\TheEvent{}
are just as consistent with an origin from a nonprecessing or precessing binary,   as long as as
the sum of the components of the spins parallel to the orbital angular
momentum $\mathbf{L}$ is nearly zero.
If the binary black hole formed from isolated stellar evolution, one could reasonably expect all angular momenta to be
strictly and positively aligned at coalescence; see \PaperAstro{} and \cite{gwastro-EventPopsynPaper-2016}.    Hence, if we believe \TheEvent{} formed from an isolated
binary,  our data would suggest black holes are born
with small spins: $a_1=|\mathbf{S}_1|/m_1^2 \le \CONSTRAINTALIGNEDSPINONELIMIT$ and $a_2 \le \CONSTRAINTALIGNEDSPINTWOLIMIT$, where $\mathbf{S}_i$ and $m_i$ are the
black hole spins and masses [\PaperPE{}].
If these strictly-aligned isolated evolution formation scenarios are true, then a low black hole spin constrains the relevant  accretion, angular momentum transport, and tidal interaction
processes in the progenitor binary; cf.
\cite{2016arXiv160103718M,gwastro-EventPopsynPaper-2016,popsyn-LowMetallicityImpact4-ConfrontUpperLimitsAndKicks2015}.  
On the other hand, the data are equally consistent with a strongly-precessing black hole binary with large component spins, formed
in a densely interacting stellar cluster (\PaperAstro).   
Measurements of the binary black holes' transverse spins will therefore provide vital clues as to the processes that
formed \TheEvent.  
In this work we use numerical relativity to  check for any  evidence for or against  spin precession that might otherwise be
obscured by model systematics.   Like \PaperPE{}, we find results consistent with but with no strong support for precessing spins.

\section{Methods}
\label{sec:Methods}

\subsection{Numerical relativity simulations of binary black hole coalescence}
\label{sec:sub:NRHistory}
The first attempts to solve the field equations of general relativity numerically
began in the 1960s, by Hahn and Lindquist \cite{1964AnPhy..29..304H}, followed with some success by
Smarr \cite{1975PhDT.........2S,1977NYASA.302..569S}.  
In the 1990s, a large collaboration of US universities
worked together to solve the ``Grand Challenge'' of evolving binary 
black holes \cite{Cook:1997na,Abrahams:1997ut,Gomez:1998uj}. 
In  2005, three groups \cite{Pretorius:2005gq, Campanelli:2005dd,Baker:2005vv} developed two completely independent techniques
that produced the first collisions of orbiting black holes. The first
technique \cite{Pretorius:2005gq} involved the use of generalized 
harmonic coordinates
and excision of the black hole horizons, while the second technique
\cite{Campanelli:2005dd,Baker:2005vv}, dubbed ``moving punctures approach'',
used singularity avoiding slices of the black hole spacetimes.

Since then, the field has seen an explosion of activity and improvements in methods and capabilities; see, e.g., \cite{2001CQGra..18R..25L,2009CQGra..26k4001H,2010RvMP...82.3069C,2014ARAA..52..661L}.
Multiple approaches have been pursued and validated against one another \cite{Baker:2007fb,Hannam:2009hh}.
Binaries can now be evolved in wide orbits \cite{Lousto:2013oza}; %
at high mass ratios up to 100:1 ~\cite{Lousto:2010ut,Sperhake:2011ik}; with near-maximal black hole spin
\cite{Lovelace:2014twa,Scheel:2014ina,Ruchlin:2014zva}; and for many orbits 
before coalescence \cite{2015PhRvL.114n1101L,Szilagyi:2015rwa}.  
At sufficiently large separations, despite small gauge and frame ambiguities,  the orbital and spin dynamics evaluated using numerical relativity agrees with
post-Newtonian calculations
\cite{2009PhRvD..79h4010C,2014PhRvD..89f1502T,2014PhRvD..89h4006P,2013PhRvL.111x1104M,2015PhRvD..92j4028O}.   
The stringent phase and amplitude needs of gravitational wave detection and parameter estimation prompted the
development of revised standards for waveform accuracy \cite{2013CQGra..31b5012H,gr-nr-WaveformErrorStandards-LBO-2008}.  
Several projects have employed numerical relativity-generated waveforms to assess gravitational-wave detection and parameter estimation
strategies \cite{Aylott:2009ya, Aylott:2009tn, Ajith:2012az, Aasi:2014tra,2014PhRvD..89d2002K,2015PhRvD..92j2001K}.  
These results have been used to calibrate models for the leading-order radiation emitted from binary black hole coalescence
\cite{2014PhRvD..89f1502T,2014PhRvD..89h4006P,nr-Jena-nonspinning-templates2007,gwastro-nr-Phenom-Lucia2010,gwastro-mergers-IMRPhenomP,2015PhRvL.115l1102B,gwastro-mergers-nr-PrayushAlignedSpinAccuracy-2016}.  
Our study builds on this past decade's  experience with modeling the observationally-relevant dynamics and radiation from
comparable-mass coalescing black holes.

In this and most NR work, the initial data for a simulation of coalescing binaries are characterized by the properties
and initial orbit of its two component black holes: by  initial black holes masses
$m_1,m_2$ and spins
$\mathbf{S}_1,\mathbf{S}_2$, specified in a quasicircular orbit such that the (coordinate) orbital angular momentum is
aligned with the $\hat{z}$ axis and the initial separation is along the $\hat{x}$ axis.  In this work, we characterize these
simulations by the dimensionless mass ratio $q=m_1/m_2$ (adopting the
convention $m_1\geq m_2$); the dimensionless spin
parameters $\boldsymbol{\chi}_i = \mathbf{S}_i/m_i^2$; and an initial dimensionless orbital frequency $M\omega_0$.  
For each simulation, the orientation-dependent  gravitational wave strain $h(t,r,\hat{n})$  at large distances can be efficiently characterized by a (spin-weighted)
spherical harmonic decomposition of $h(t,r,\hat{n})$ as $h(t,r,\hat{n}) = \sum_{l\ge 2}\sum_{m=-l}^l h_{lm}(t,r)
\Y{-2}_{lm}(\hat{n})$.   To a good first approximation, only a few terms in this series are necessary to characterize the
observationally-accessible radiation in any direction
\cite{2007PhRvD..76h4020V,gwastro-spins-rangefit2010,2013PhRvD..88b4034H,2014PhRvD..90l4004V,2016PhRvD..93h4019C}.     
For example,  when a binary is widely separated, only two terms dominate this sum: $(l,|m|)=(2,2)$.  Conversely, however,
several terms (modes) are required for even nonprecessing binaries, viewed along a generic line of sight; more are
needed to capture the radiation from precessing binaries. 

For nonprecessing
sources with a well-defined axis of symmetry, 
individual modes $(l,m)$ 
have distinctive characters, and can be easily isolated
numerically and compared with analytic predictions.  For precessing
sources, however, rotation mixes modes with the same $l$.
To apply our procedure self-consistently to both
nonprecessing and precessing sources, we include all modes $(l,m)$ with the same $l$.    
However, at the start of each simulation, the $(l,m)$ mode oscillates at  $m$ times the orbital frequency.  For $m>3$,
scaling our simulations to the inferred mass of the source,  this initial mode frequency is often well above
$30\,\unit{Hz}$, the minimum frequency we adopt in this work for parameter estimation.  We therefore cannot safely and
self-consistently compare all modes with $l>3$ to the data using numerical relativity alone: an  approximation would be required
 to go to higher order (i.e., hybridizing each NR simulation with an analytic approximation at early times).

Therefore,  in this paper, we use all five of the $l=2$
modes to draw conclusions about \TheEvent{}, necessary and sufficient to capture the leading-order influence of any
orbital precession.    To incorporate the effect of higher harmonics, we repeat our calculations, using
  all of the $l\le 3$ modes.    We defer a careful treatment of higher-order modes and the $m=0$ modes to \PaperPENRMethods{}
and subsequent work.

\subsection{Numerical relativity simulations used}
\label{sec:Methods:NR:Sims}

Our study makes use of \SimulationsProvidedNumber{} distinct simulations of binary black hole quasicircular inspiral and
coalescence.   Table \ref{tab:SimulationList} summarizes the salient features of this set: mass ratio and initial spins for the simulations used here, all initially in a
quasicircular orbit with orbital separation along the $\hat{x}$ axis and velocities along $\pm \hat{y}$.  

The RIT group provided \SimulationsProvidedNumberRIT{} simulations
\cite{2014PhRvD..90j4004H,2015PhRvD..92b4022Z,2015PhRvL.114n1101L}.  
The simulations include binaries with a wide range of mass ratios, as well as a wide range of black hole angular
momentum (spin) magnitudes and directions
\cite{2015PhRvL.114n1101L,2014PhRvD..90j4004H,2015PhRvD..92b4022Z,2013PhRvD..87h4027L}, 
including a  simulation with large transverse spins and several spin precession cycles which fits \TheEvent{} well
\cite{2015PhRvL.114n1101L}, as described below.
The SXS group has provided both a publicly-available catalog of coalescing black hole binary mergers
\cite{gr-nr-CornellCaltechCatalogDump-2013}, a new catalog of nonprecessing simulations \cite{2015arXiv151206800C},  and selected supplementary simulations described below.  Currently extended to \SimulationsProvidedNumberSXS{} members in the form
used here, this catalog includes  many high-precision zero-
and  aligned-spin sources; selected precessing systems; and simulations including extremely high black hole spin.
The Georgia Tech group (GT) provided \SimulationsProvidedNumberGT{} simulations; see  
 \cite{gwastro-mergers-nr-Alignment-ROS-CorotatingWaveforms} and \cite{gr-nr-vacuum-GTcatalog-2016} for further details.  
This extensive archive  covers a wide range of spin magnitudes and orientations, including several systematic one- and
two-parameter families. 
The Cardiff-UIB group provided \SimulationsProvidedNumberBAM{} simulations, all specifically produced to follow up
\TheEvent{} via  a high-dimensional grid stencil, performed via the BAM code \cite{PhysRevD.77.024027,husa2008}.  
These four sets of simulations explore the model space near the event in a well-controlled fashion.
In addition to previously-reported simulations, several groups performed new simulations (\SimulationsProvidedNumberFollowupTotal{} in total) designed to reproduce the
parameters of the event, some of which were applied to our analysis.  These simulations are denoted in
Table \ref{tab:SimulationList} and our other reports by an asterisk (*).  
These followup simulations include three independent simulations of
the same parameters drawn from the distributions in \PaperPE{}, from RIT, SXS, and GT, allowing us to assess our
systematic error.  These simulations were reported in  \PaperDetection{} and \PaperBurst{}, and are indicated by (+) in our tables.

The simulations used here have either been published previously, or were produced using one of three well-tested procedures
operating in familiar circumstances. 
For reference, in Appendix \ref{ap:FullSimulationDetails}, we outline the three groups' previously-established methods
and results.  For this application,  we trust these simulations' accuracy, based on their past track record of good
performance.  By incorporating simulations of identical physics provided by different groups, our methods provide
limited direct corroboration: simulations with similar physics produce similar results.   

\subsection
    {\label{sec:Methods:PE}Directly comparing NR with data}

For each simulation, each choice of seven extrinsic parameters $\theta$ (4 spacetime coordinates for the coalescence
event; three Euler
angles for the binary's orientation relative to the Earth), and each choice for the redshifted total binary mass $M_z=(1+z) M$, we can predict the response $h_k$ of both
of the $k=1,2$ LIGO
instruments to the implied gravitational wave signal.   Using $\lambda$ to denote    the combination of  redshifted mass $M_z$ and
the numerical relativity simulation parameters needed to uniquely specify the binary's dynamics,  we can therefore
evaluate the likelihood of the data given the noise:
\begin{align}
\ln {\cal L}(\lambda; \theta) 
&= -\frac{1}{2}\sum_k \qmstateproduct{h_k(\lambda,\theta)-d_k}{h_k(\lambda,\theta)-d_k}_k  %
 - \qmstateproduct{d_k}{d_k}_k,
\end{align}
where $h_k$ are the predicted response of the $k$th detector due to a source with parameters $\lambda,\theta$; $d_k$ are the
detector data in instrument $k$; and $\qmstateproduct{a}{b}_k \equiv \int_{-\infty}^{\infty} 2 df
\tilde{a}(f)^*\tilde{b}(f)/S_{h,k}(|f|)$ is an inner product implied by the $k$th's detector's noise power spectrum
$S_{h,k}(f)$;  see, e.g., \cite{gwastro-PE-AlternativeArchitectures} for more details.  In practice, as discussed in the
next section, we adopt a low-frequency cutoff $f_{\rm low}$, so all inner products are modified to
\begin{eqnarray}
\qmstateproduct{a}{b}_k\equiv 2 \int_{|f|>f_{\rm low}}  df
\frac{\tilde{a}(f)^*\tilde{b}(f)}{S_{h,k}(|f|)}.
\end{eqnarray}
Except for an overall normalization constant and a different choice for low-frequency cutoff, our expression
agrees with Eq. (1) in \PaperPE{}.   The joint posterior probability of $\lambda,\theta$ follows from Bayes theorem:
\begin{eqnarray}
p_{\rm post}(\lambda,\theta) = \frac{{\cal L}(\lambda,\theta) p(\theta) p(\lambda)}{ \int d\lambda d\theta {\cal
  L}(\lambda,\theta) p(\lambda)p(\theta)} \; ,
\end{eqnarray}
where $p(\theta)$ and $p(\lambda)$ are priors on the (independent) variables $\theta,\lambda$.\footnote{For simplicity,
  we assume all black hole-black hole (BH-BH) binaries are equally likely anywhere in the universe, at any orientation relative to the detector. Future direct observations may favor a
  correlated distribution, including  BH
  formation at higher masses at large redshift.}
For each $\lambda$ --- that is, for each simulation and each  redshifted mass $M_z$ --- we evaluate  the marginalized likelihood
\begin{eqnarray}
{\cal L}_{\rm marg}(\lambda) \equiv \int {\cal L}(\lambda,\theta)p(\theta) d\theta
\end{eqnarray}
via direct Monte Carlo integration, where $p(\theta)$ is  uniform in 4-volume and source orientation
\cite{gwastro-PE-AlternativeArchitectures}.\footnote{Our choice for $p(\theta)$ differs only superficially from  that adopted
in \PaperPE,  by adopting a narrower prior on the geocentric event time.  Here, we allow $\pm 0.05\, \unit{s}$ around the
time reported by the online analysis; 
 \PaperPE{} allowed $\pm 0.1\unit{s}$.  
}  
The marginalized likelihood  measures the similarity between the data and a source with parameters
$\lambda$ and enters naturally into full Bayesian posterior calculations.   In terms of the
marginalized likelihood and some assumed prior $p(\lambda)$ on intrinsic parameters like masses and spins, the posterior
distribution for intrinsic parameters is
\begin{eqnarray}
\label{eq:Posterior}
p_{\rm post}(\lambda) = \frac{{\cal L}_{\rm marg}(\lambda) p(\lambda)}{\int d\lambda {\cal L}_{\rm marg}(\lambda) p(\lambda)}.
\end{eqnarray}
If we can evaluate ${\cal L}_{\rm marg}$ on a sufficiently dense grid of intrinsic
parameters, Eq.~(\ref{eq:Posterior}) implies that
we can reconstruct the full posterior parameter distribution via interpolation or other local
approximations.    
This reconstruction needs to accurately reproduce ${\cal L}_{\rm marg}$ only
near its peak value; for example, if ${\cal
  L}_{\rm marg}(\lambda)$ can be approximated by a $d$-dimensional Gaussian, then we anticipate only configurations
$\lambda$ with 
\begin{eqnarray}
\label{eq:DefineGrayBlackCutoffs}
\ln
{\cal L}_{\rm max}/{\cal L}_{\rm marg}(\lambda) > \chi^2_{d,\epsilon}/2
\end{eqnarray}
 contribute to the posterior distribution at the
$1-\epsilon$ credible interval, where $\chi^2_{d,\epsilon}$ 
is the inverse-$\chi^2$ distribution.   

Based on similarity of our distribution to a suitably-parameterized multidimensional Gaussian,  we
anticipate 
that only the region of parameter space with 
$\ln{\cal L}_{\rm max}-\lnLmarg(\lambda)\lesssim 6.7$ 
can potentially impact our conclusions
regarding the \pCI{} credible level for $d=8$ 
(i.e., two masses and two precessing spins); for $d=4$, more relevant to
the most strongly accessible parameters (i.e., two masses and two aligned 
spins), the corresponding interval is 
$\ln{\cal L}_{\rm max}-\lnLmarg(\lambda)\lesssim 4$.

Each NR simulation corresponds to a particular value of
seven of the intrinsic parameters (mass ratio and the three
components of each spin vector) but can be scaled to an arbitrary
value of the total redshifted mass $M_z$.  Therefore each NR
simulation represents a one-parameter family of points in the 8-dimensional
parameter space of all possible values of $\lambda$.   
For each simulation, we evaluate the marginalized log likelihood versus redshifted mass $\lnLmarg(M_z)$ on an array of masses, adaptively exploring each one-parameter family to cover the interval 
$\ln{\cal L}_{\rm max}-\lnLmarg(\lambda)< 10$.
To avoid  systematic bias introduced by interpolation or fitting, our principal results are simply these tabulated function
values, explored almost continuously in mass $M_z$ and 
discretely,  as our fixed simulation archive permits,
 in other parameters.    The set of
intrinsic parameters $V_C\equiv \{\lambda : \lnLmarg > C\}$ above a cutoff $C$ identifies a
subset of binary configurations whose gravitational wave emission is consistent with the data.\footnote{While this approach works for multidimensional Gaussians, it can break
  down in coordinate systems where the prior is particularly significant (e.g., diverges; see the grid-based
  method in \cite{2015CQGra..32w5017H}) or where the likelihood has strong features (e.g., corners and tails) in
  multiple dimensions.  
For example, a likelihood constant on a sphere plus thin, long spines (e.g., the  shape of a sea urchin)  will have little posterior support on the spines, but
  each of the spines would be selected by a likelihood cut of the kind used here.  As our calculations below demonstrate, marginalization over extrinsic parameters eliminates most
  complexity in the likelihood: our function is smooth, dominated by a handful of parameters, without corners or narrow tails.  }
Though this approach provides a powerfully model-independent approach to gravitational-wave parameter estimation, as
described above it is restricted to the discrete grid of NR simulation values.  Fortunately, the brevity and simplicity
of the signal --- only a few chirping and little-modulated cycles --- requires the posterior distribution to be broad and
smooth, extending over many numerical relativity simulations' parameters.
This allows us to go beyond comparisons on a discrete grid of NR simulations,
and instead interpolate between simulations to reconstruct the entire 
distribution.

\begin{figure}
\includegraphics[width=\columnwidth]{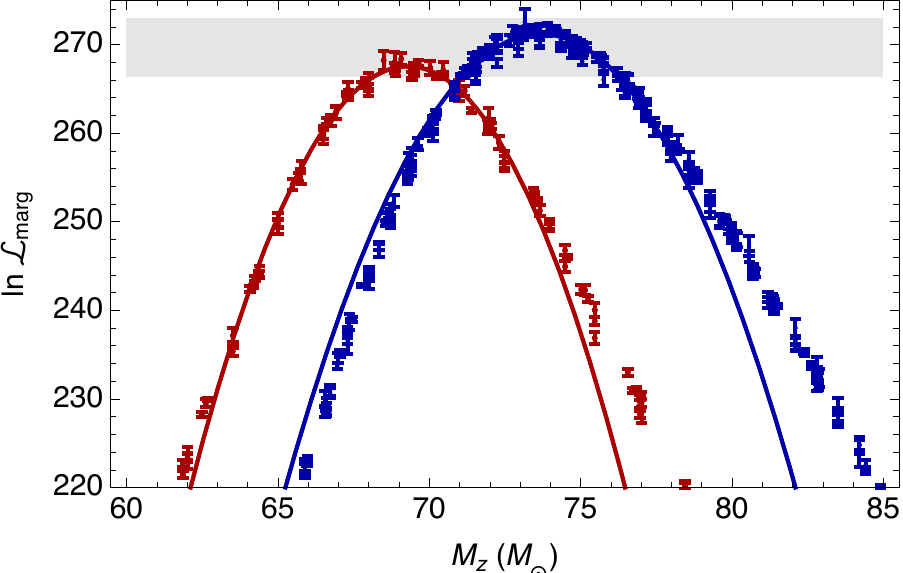}
\caption{\label{fig:lnL:VersusMass}\textbf{Likelihood versus mass: Examples}: Raw Monte Carlo estimates for
  $\lnLmarg(M_z)$ versus $M_z$ for two nonprecessing binaries: SXS:BBH:305  (blue)  and
  \texttt{d0\_D10.52\_q1.3333\_a-0.25\_n100} (red).   
To guide the
  eye, for each simulation we also overplot a local quadratic fit to the results near each peak.    Results were
  evaluated with $f_{\rm min}=30\,\unit{Hz}$; compare to Table \ref{tab:SimulationRanks:30}.  Error bars reflect
  the standard Monte Carlo estimate of the integral standard deviation, multiplied by $2.57$ in the log to increase
  contrast (i.e., the nominal 99\% credible interval, assuming the relative Monte Carlo errors are normally distributed).
To guide
  the eye, a shaded region indicates the interval of $\lnLmarg$ selected
  by our ansatz given a credible interval \pCI{} and a peak value of $\lnLmarg$ of  273; see Section
  \ref{sec:Methods} and Eq. (\ref{eq:DefineGrayBlackCutoffs}).
}
\end{figure}

To establish a sense of scale, we can use a simple order-of-magnitude calculation for
$\lnLmarg$.  The signal to noise ratio  $\rho$ and peak likelihood of any assumed signal are related:  $\rho=\sqrt{2\text{max}_{\theta} \ln {\cal L}}$.  
   Even at the best intrinsic parameters $\lambda$, the marginalized log-likelihood
  $\lnLmarg$ will be well below the peak value $\text{max}_{\theta} \ln {\cal L}$,  because only a small fraction of extrinsic parameters $\theta$ have support
  from the data \cite{gwastro-mergers-HeeSuk-CompareToPE-Aligned}.  Using \TheEvent{}'s previously-reported signal amplitude
  [$\rho = \PEOPTIMALSNRRANGE$],   its extrinsic parameters and their uncertainty [\PaperPE],  and our prior $p(\theta)$,  we
  expect the peak value of $\lnLmarg$ to be of order 240-330.
  The interval of $\lnLmarg$ selected by  Eq. (\ref{eq:DefineGrayBlackCutoffs}) is a small fraction of the full range of
  $\lnLmarg$, identifying a narrow range of parameters $\lambda$ which are consistent with the data.
Our analysis of this event, as well as synthetic data, suggests that $\lnLmarg$ is often
well-approximated by simple low-order series in intrinsic parameters $\lambda$.   This simple behavior is most apparent
versus total mass $M_z$.  
Figure  \ref{fig:lnL:VersusMass} shows examples of the marginalized log likelihood evaluated using two
of our most promising simulation candidates: they
are well-approximated by a quadratic over the entire
observationally-interesting range. 
We approximate $\lnLmarg(M_z)$ as a second-order Taylor series,
\begin{eqnarray}
\label{eq:lnL}
\lnLmarg(M_z) \simeq  \ln L  -  \frac{1}{2} \Gamma_{MM}(M_z-M_{z,*})^2,
\end{eqnarray}
where the constants $\ln L$, $M_{z,*}$, and $\Gamma_{MM}$ represent
the largest value of $\lnLmarg$, the redshifted mass at which this
maximum occurs, and the second derivative at the peak value.
Even in (rare) cases when a locally quadratic approximation slightly breaks down, we still use $\ln L$ to denote our estimate of the peak
  of $\lnLmarg(M_z)$.\footnote{We find similar results using more sophisticated nonparametric interpolation schemes.
    The results reported in Table \ref{tab:SimulationRanks:30} use one-dimensional Gaussian process interpolation to
    determine the peak value.
}   As a means of efficiently communicating trends in the quality of fit versus intrinsic parameters, the two
  quantities $\ln L$ and $M_{z,*}$ are reported in Table \ref{tab:SimulationRanks:30}.   
Motivated by the success of this approximation,  in Section \ref{sec:sub:aligned} we also supply a quadratic
approximation to $\lnLmarg$ near its peak, \emph{under the restrictive approximation that all angular momenta are
  parallel}, using information from only nonprecessing simulations.  Using this quadratic approximation, we can numerically
estimate $\lnLmarg$ and hence the posterior [Eq. (\ref{eq:Posterior})] for arbitrary aligned-spin binaries.  For any
coordinate transformation
$z=Z(\lambda)$,  we can use suitable supplementary coordinates and direct numerical quadrature to determine the marginal posterior density $p_{\rm post}(z)  = \int p_{\rm post}(\lambda)
\delta(z-Z(\lambda))$.   As shown below, this procedure yields results comparable to \PaperPE{} for nonprecessing
binaries.

\subsection{Are there sufficiently many and long NR simulations?}
\label{sec:NR:DurationIssues}

Because of finite computational resources, NR simulations 
of binary black holes
cannot include an arbitrary number of orbits before merger. Instead, they
start at some finite initial orbital frequency.
While many NR simulations 
follow enough binary orbits to 
be compared with \TheEvent{} over the entire LIGO frequency band, some NR
simulations miss some early-time information.  
Therefore, in this section
we  describe a simple approximation 
(a low frequency cutoff) 
we apply to ensure that simulations with similar physics
(but different initial orbital frequencies) 
lead to similar results.

At the time of \TheEvent{}, the instruments had relatively poor sensitivity to frequencies below $30\unit{Hz}$ and
almost no sensitivity below $20 \unit{Hz}$.  For this reason, the interpretations adopted in \PaperPE{} adopted a
low-frequency cutoff of $20\unit{Hz}$.  Because of the large number of cycles accumulated at low
frequencies, a straightforward Fisher matrix estimate
\cite{1995PhRvD..52..848P,gw-astro-mergers-NRParameterEstimation-Nonspinning-Ajith}
 suggests these low frequencies ($20-30\unit{Hz}$) provide
a nontrivial amount of information, particularly about the binary's  total mass.   Equivalently, using the techniques described in this
paper, the function $\lnLmarg(M_z)$ will have a slightly higher and narrower peak when including 
all frequencies
than when truncating the signal to only include frequencies above $30\unit{Hz}$; see \PaperPENRMethods{}.  

\begin{figure}
\includegraphics[width=\columnwidth]{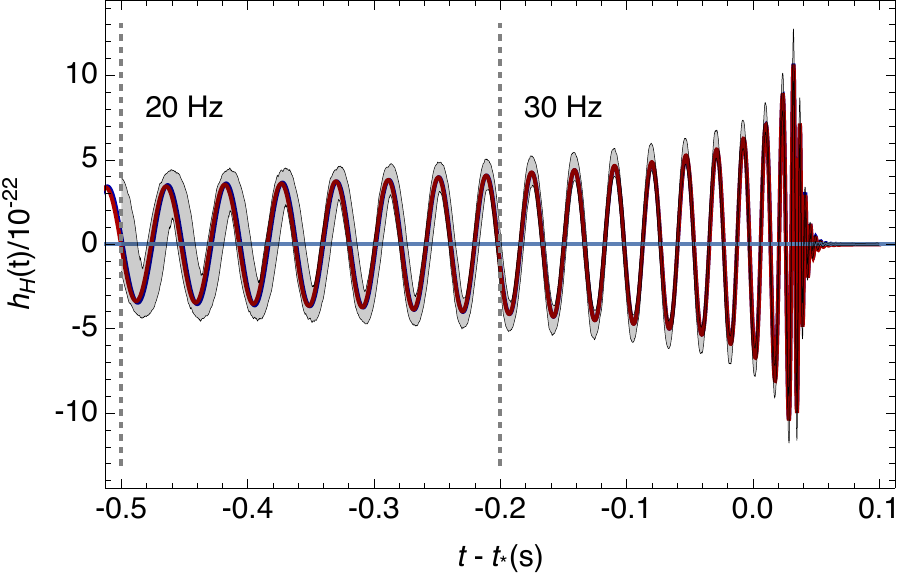}
\caption{\label{fig:StrainExamples}\textbf{Best-fit detector response}: A plot of the detector response (strain) 
  $h(t)=F_+h_+(t)+F_\times h_\times(t)$ evaluated at the LIGO-Hanford detector , similar to  Figure 2 in
    \PaperDetection{}, Figure 6 in \PaperPE, and Figure 2 in \PaperTestGR, evaluated using two of the
    best-fitting  numerical
    relativity simulations and total redshifted masses  reported in Table  \ref{tab:SimulationRanks:30}.  The redshifted masses $M_z$ and
    extrinsic parameters $\theta$  necessary to evaluate the detector response have  been identified using the methods
    used in this work and \PaperPENRMethods, using all $l=2$ modes; a low-frequency cutoff of 30 Hz; and omitting the impact of
    calibration uncertainty.  For comparison, the shaded region shows the 95\% credible region for the
    waveform reported in \PaperPE{}, an analysis  which accounts for calibration uncertainties and includes frequencies
    down to 20 Hz but  approximates the radiation and omits higher harmonics (e.g, the $(2,\pm 1)$ modes).  To guide the eye, two vertical lines indicate the approximate time at which the signal crosses these
    two gravitational wave frequency thresholds. 
  }
\end{figure}

Because of limited computational resources, relatively few simulations start in a sufficiently wide orbit such that, for
$M_z=70 M_\odot$, their radiation in the most significant harmonics of the orbital frequency will be at or below the
lowest frequency ($20 \unit{Hz}$) adopted in \PaperPE{}.    If $f_{\rm min}$ is the  low-frequency cutoff,
$M\omega_0 /m \lesssim  0.02 (M_z/70 M_\odot) (f_{\rm min}/20\unit{Hz})(2/m)$,
where  $M\omega_0$ is the initial orbital frequency of the simulation
reported 
in Table \ref{tab:SimulationList},
can be safely used
to analyze a signal containing a significant contribution from the $m$th harmonic of the orbital frequency.  Figure
\ref{fig:StrainExamples} shows examples of the strain in LIGO-Hanford, predicted using simulations of different
intrinsic duration, superimposed with lines approximately corresponding to different gravitational wave frequencies.
To facilitate an apples-to-apples comparison incorporating the widest range of available simulations, in this work we
principally report on comparisons calculated by adopting a low-frequency cutoff of $30\unit{Hz}$; see, e.g., Table
\ref{tab:SimulationRanks:30}.   (We also briefly report on comparisons performed using a low-frequency cutoff of $10\unit{Hz}$.)
As we describe in subsequent sections, while this choice of $30\unit{Hz}$ slightly degrades our ability to make subtle distinctions between different
precessing configurations, it does not dramatically impair our ability to reconstruct parameters of the event, given
other significant degeneracies.  

Even this generous low-frequency cutoff is not perfectly safe:  for each simulation, a minimum mass exists at which the starting
gravitational wave frequency is $30\unit{Hz}$ or larger.  
In the plots and numerical results reported here,  
we have eliminated simulation and mass
choices that correspond to scaling an NR simulation to a starting 
frequency above $30\unit{Hz}$.  The inclusion or suppression of
these configurations does not significantly change our principal results.

This paper uses enough NR simulations to adequately sample the four-dimensional space of nonprecessing spins,
particularly for comparable masses.  
As described below, this high simulation density insures we can
reliably approximate the marginalized likelihood $\lnLmarg$ for nonprecessing systems.  
 On the other hand, the eight-dimensional parameter space of precessing binaries is much more sparsely explored by the
 simulations available to us.  
But because the reconstructed gravitational wave signals 
in \PaperDetection{} and \PaperPE{} exhibit little to no modulation,
we expect that the remaining four parameters 
must have at best a subtle effect on
the signal: the likelihood and posterior distribution should depend only weakly on any additional subdominant parameters.
Having identified dominant trends  using nonprecessing simulations, we can use 
 controlled sequences of simulations with similar parameters to determine the residual impact of transverse spins.
 Even if the marginalized
likelihood cannot be safely approximated in general, a simulation's value of $\ln L$ provides insight into the
parameters of the event.

Motivated by the parameters reported in \PaperPE{} and our results in Table
\ref{tab:SimulationRanks:30}, several followup simulations were performed to reproduce \TheEvent{}.   These simulations
are responsible for most of the best-fitting aligned-spin results reported in  Table \ref{tab:SimulationRanks:30}.

\subsection{Impact of instrumental uncertainty}
For simplicity, our analysis does not automatically account for instrumental
uncertainty (i.e., in the detector noise power spectrum or instrument calibration), as do the methods in \PaperPE{}.  
 \PaperPE{}
suggests that,  for the intrinsic parameters $\lambda$ of interest here,  the impact of these systematic instrumental uncertainties effects are relatively small.  We have  repeated our analysis using two
versions of the instrumental calibration; we find no significant change in our results.

\subsection{Comparison with other methods}
\label{sec:Compare}
\PaperBurst{} reported on direct comparisons between radiation extracted from NR simulations and \emph{nonparametrically
  reconstructed} estimates of the gravitational wave signal; see, e.g., their Fig.~12.  Their comparisons quickly identified masses, mass
ratios, and spins that were consistent with the data.
Our study, which attempts a fully Bayesian
direct comparison between the data and the multimodal predictions of NR,
produces results consistent with those of \PaperBurst{}; 
see, e.g., Figure~\ref{fig:Constraints30:MQXi} described below.

\PaperPE{} performed Bayesian inference on the data using semianalytic
models for the gravitational waves from a coalescing compact binary.
We directly compare our posterior distribution with that of \PaperPE{}
for the special case of aligned spins.
Differences between these posterior distributions can be due to many factors:
our choice of starting
frequency is slightly higher ($30\unit{Hz}$ versus $20\unit{Hz}$);  our approach does not account for calibration uncertainty; and of course we employ
NR instead of a semianalytic waveform model.
To isolate the effects of NR, we have repeated our analysis
but with the same nonprecessing waveform model used in \PaperPE{} rather than
with NR waveforms.  Using the same input waveforms, our method and that
of \PaperPE{} produce very similar results; 
see \PaperPENRMethods{} for details.
To isolate the effects of the low-frequency cutoff, we performed the nonprecessing analysis reported in  \PaperPE{} with a low frequency cutoff of
$30\unit{Hz}$; we found results similar to \PaperPE{}.   

\section{\label{sec:Results}Results I: Pre-coalescence parameters}

We present two types of results.  For generic, precessing NR simulations, we 
evaluate the marginalized likelihood of source parameters given the data,
but because the parameter-space coverage of NR simulations is so sparse,
we do not attempt to construct an interpolant for the likelihood as
a function of source parameters.  For nonprecessing sources, we construct
such an interpolant, and we compare with the results of \PaperPE{}.
Using the computed likelihoods,
we quantify whether the data
are consistent with or favor a precessing source.

\begin{table*}
\begin{ruledtabular}
\begin{tabular}{l|cccc|ccc}
- & NR grid & Aligned fit & Overall  & LI & NR grid ($l\le 3$) & Aligned fit ($l\le 3$) & Overall ($l\le 3$) \\ \hline
Detector-frame initial total mass $M_z (M_\odot)$   & 65.6--77.7 & 67.2 -- 77.2 &  65--77.7 & \MTOTobsRANGE  & 67.1--76.8   & 67.2.3--77.3 & 67.1--77.3  \\
Detector-frame  $m_{1,z} (M_\odot)$   & 35--45  & 35--45 & 35--45   & \MONEobsRANGE  & 34.5--43.9 & 35--45 & 34.5--45 \\
Detector-frame  $m_{2,z} (M_\odot)$   & 27--36  & 27--36.7 & 27--36.7  & \MTWOobsRANGE & 30--37.5 & 28--37 & 28--37.5 \\
Mass ratio $1/q$  & 0.66--1 & 0.62--1 & 0.62--1 &  \MASSRATIORANGE & 0.67--1 & 0.69--1 & 0.67--1\\  %
Effective spin  $\chi_{\rm eff}$ & -0.3 -- 0.2 & -0.2 -- 0.1 & -0.3--0.2  &  \CHIEFFRANGE & -0.24 -- 0.1 & -0.2--0.1 & -0.24--0.1\\
Spin 1 $a_1$ &0--0.8  & 0.03--0.80 & 0--0.8 &  \SPINONERANGE & 0--0.8 & 0.03--0.83 & 0--0.83 \\
Spin 2 $a_2$ &0--0.8 & 0.07--0.91 & 0--0.91  & \SPINTWORANGE & 0--0.8 & 0.11-- 0.92 & 0--0.92  \\
\hline
Final total mass $M_{f,z} (M_\odot)$  & 64.0--73.5 & - & \ReportedNumberThisPaperMfzMsunLow --
\ReportedNumberThisPaperMfzMsunHigh & \MFINALobsRANGE & 64.2--72.9 & & 64.2--72.9\\
Final spin $a_f$ & 0.62--0.73 &   &  \ReportedNumberThisPaperAFLow-- \ReportedNumberThisPaperAFHigh & \SPINFINALRANGE &
0.62--0.73 & & 0.62--0.73
\end{tabular}
\end{ruledtabular}
\caption{\label{tab:ParameterRanges}\textbf{Constraints on $M_z,q,\chi_{\rm eff}$}: Constraints on  selected parameters of \TheEvent{} derived by directly comparing
  the data to numerical relativity simulations.  The first column reports the extreme values of each parameter consistent
  with   $\lnLmarg>\lnLmargBlack$ [Eq. (\ref{eq:DefineGrayBlackCutoffs}), with $d=4$],  corresponding to
the black points shown in Figures
  \ref{fig:Constraints30:MQXi}, \ref{fig:Constraints30:Chi1zChi2z}, and \ref{fig:FinalState}.  These are computed using all the $l=2$ modes of the NR
waveforms.
Because these extreme values are evaluated only on a sparse discrete grid of NR simulations, this procedure can underestimate the extent of the allowed range of each
parameter.
The second column reports the \pCI{} credible interval derived by fitting $\lnLmarg$ versus these parameters for
nonprecessing binaries, to enable 
interpolation between points on the discrete grid in $\lambda$; see Section \ref{sec:sub:aligned} for details.  
The third column is the union of the two intervals.  For comparison, the fourth column provides the interval reported in
\PaperPE{}, including precession and systematics.  The remaining three columns show our results derived using all $l\le
3$ modes.
}
\end{table*}

\subsection{Results for generic sources, without interpolation}

Because the available generic NR simulations represent only a sparse sampling
of the parameter space, for generic sources
we adopt a conservative approach: we rely only on our estimates
of the marginalized likelihood $\lnLmarg$, 
and we do not interpolate the likelihood between intrinsic parameters, nor
do we account for Monte Carlo
uncertainty in each numerical estimate of $\lnLmarg$.
Using the inverse $\chi^2$ distribution, we identify two thresholds in $\lnLmarg$ using
Eq. (\ref{eq:DefineGrayBlackCutoffs}), one (our preferred choice) obtained by adopting  $d=4$
observationally accessible parameters, and the other adopting
$d=8$.\footnote{The second choice ($d=8$) would be appropriate if the 
posterior was well-approximated by an 8-dimensional Gaussian.  
The first  choice ($d=4$) is motivated by past parameter estimation studies when
the posterior distribution principally constrains the component masses and
aligned spins.  
}
Both thresholds on $\lnLmarg$ are derived using (a) our target credible
interval (\pCI) and (b) the peak log likelihood attained over all simulations 
[Table \ref{tab:SimulationRanks:30}].  
Below, we find that the peak log likelihood over all simulations is 
$\lnLmarg=\lnLmargMax$; as a result,  these two thresholds
are $\lnLmarg=\lnLmargBlack$ and $\lnLmarg=\lnLmargGray$, for
$d=4$ and $d=8$, respectively.
The configurations of
masses and intrinsic parameters that pass either of
these two thresholds 
are deemed consistent with the data.
In subsequent figures, we will color these two classes of configurations 
in black (those configurations with $\lnLmarg>\lnLmargBlack$)
and 
gray (those configurations with $\lnLmarg>\lnLmargGray$). 
We use this set
of points in parameter space
to bound (below) the range of parameters consistent with the data.     

\begin{figure*}
\includegraphics[width=\columnwidth]{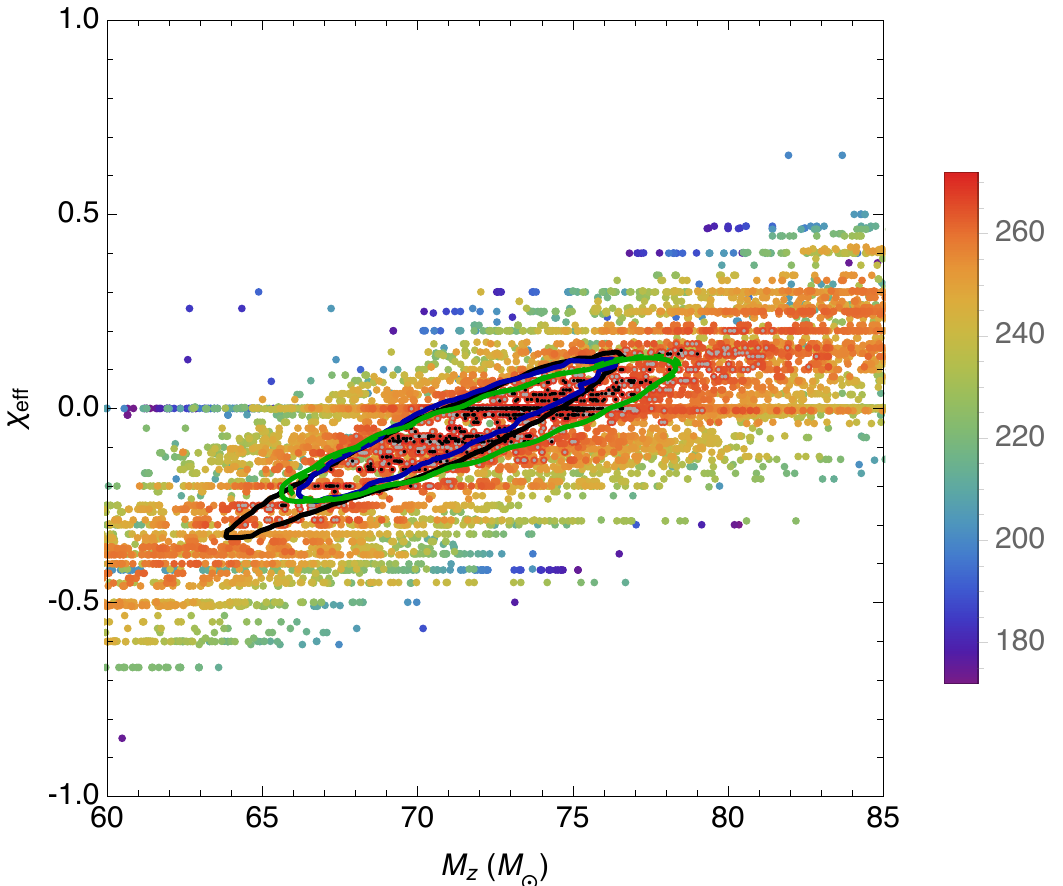}
\includegraphics[width=\columnwidth]{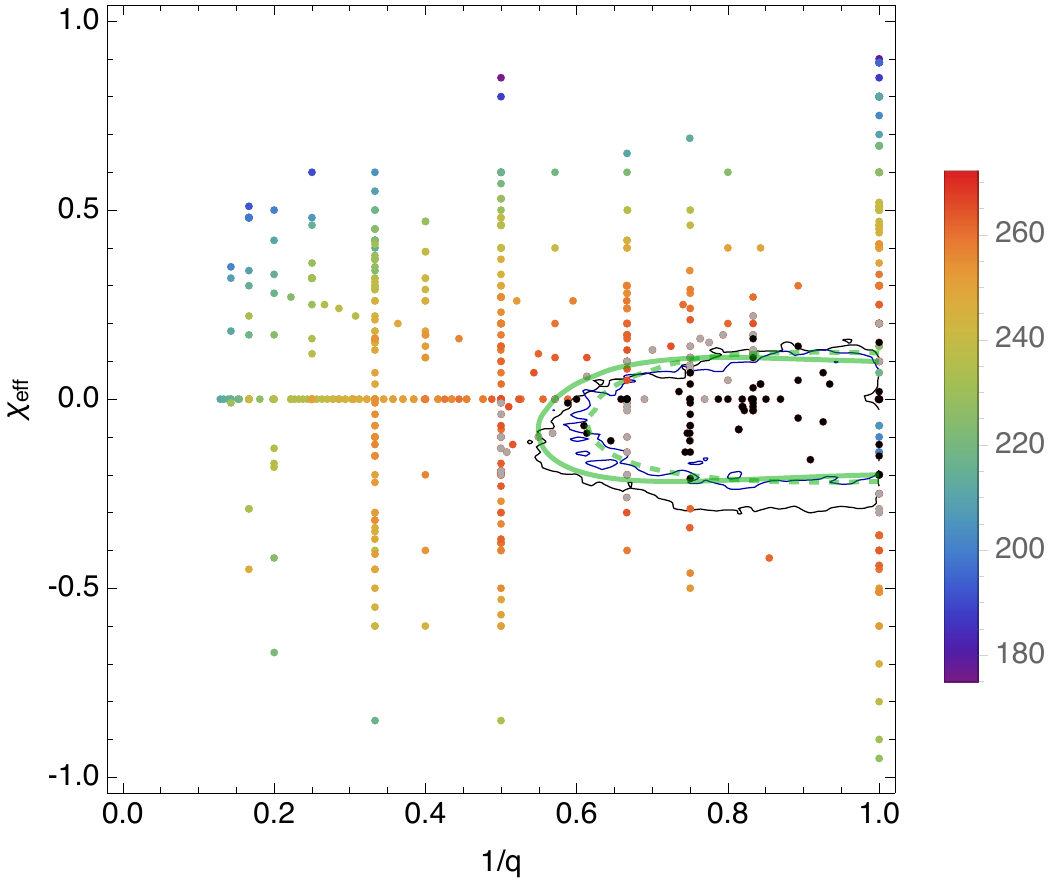}
\caption{\label{fig:Constraints30:MQXi}
\textbf{Mass, mass ratio, and effective spin are constrained and correlated}:
 Colors represent the marginalized log likelihood as a function
    of
  redshifted total mass $M_z$, mass ratio $q$ and effective
  spin parameter $\chi_{\rm eff}$.
  Each point represents an NR simulation and a particular $M_z$.
  Points with
    $\lnLmargGray < \lnLmarg < \lnLmargBlack$ are shown in light gray,
  with $\lnLmarg > \lnLmargBlack$ are shown in black, and 
  with $\lnLmarg < \lnLmargGray$ are shown according to the
  color scale on the right (points with $\lnLmarg< 172$
  have been suppressed to increase contrast).
Marginalized likelihoods are computed
  using $f_{\rm low}=30\unit{Hz}$, 
using all $l=2$ modes, and without correcting for
  (small) Monte Carlo integral uncertainties.
These figures include both nonprecessing and precessing simulations.
 For
  comparison, the \TheColorSEOB{}, \TheColorIMRP{}, and green contours show estimated 
\pCI{} credible intervals, calculated assuming that 
  the binary's spins and orbital
  angular momentum are parallel.  The solid \TheColorSEOB{} contour corresponds to the \pCI{} credible interval reported in
  \PaperPE{}, assuming spin-orbit alignment; the solid \TheColorIMRP{} 
  contour shows the corresponding \pCI{} interval reported using
  the semianalytic precessing model (IMRP) in \PaperPE{}; the solid green curve shows the \pCI{} credible intervals derived using
  a quadratic fit to $\lnLmarg$ for nonprecessing simulations using $l=2$ modes; and the dashed green curve shows the \pCI{} credible intervals
  derived using $\lnLmarg$ from nonprecessing simulations, calculated using all modes with $l\le 3$; see Section \ref{sec:sub:aligned} for details.    
  Unlike our calculations,  the \TheColorSEOB{} and \TheColorIMRP{} contours
from \PaperPE{}
  account for calibration uncertainty and use a low frequency cutoff of 20 Hz.
\emph{Left panel}: Comparison for $M_z,\chi_{\rm eff}$.   This figure demonstrates the strong correlation between the total
redshifted mass and spin.  %
\emph{Right panel}: Comparison for $q,\chi_{\rm eff}$.  
This figure is consistent with the similar but simpler analysis reported in \PaperBurst{};   see, e.g., their
Fig. 12.
}
\end{figure*}

For the progenitor black hole
parameters, our results using $l=2$ modes are summarized in Figures 
\ref{fig:Constraints30:MQXi} and \ref{fig:SpinDiskPlot}  (for generic sources), as
well as by Figures \ref{fig:ap:AlignedSpinMaxLmarg} and  \ref{fig:Constraints30:Chi1zChi2z} (for nonprecessing sources).  For comparison, these figures also include the results obtained in \PaperPE{}, 
using approximations appropriate for nonprecessing 
(\TheColorSEOB{} curves) and simply \cite{ACST} precessing 
(\TheColorIMRP{} curves) binaries.  
The first column of Table \ref{tab:ParameterRanges} shows the one-dimensional range inferred for each parameter by our
threshold-based method,  using $l=2$ modes only.  

Before describing our results, we first 
demonstrate why our strategy is
effective: Figures~\ref{fig:Constraints30:MQXi}, \ref{fig:ap:AlignedSpinMaxLmarg}, \ref{fig:Constraints30:Chi1zChi2z}, and \ref{fig:SpinDiskPlot} show
that the likelihood is smooth and slowly varying, dominated by a few key parameters.  
As seen in the right panel of Figure \ref{fig:Constraints30:MQXi}, even our large NR array is relatively sparse.
However, as the color scale on this and other figures indicate, the marginalized likelihood varies smoothly with parameters,
over a range of more than $e^{100}$.  The simplicity of $\lnLmarg$ is most apparent using controlled one- and two-parameter
subspaces; for example, Figure \ref{fig:ap:AlignedSpinMaxLmarg} shows 
that $\ln L$ (i.e., the peak of $\lnLmarg(M_z)$) varies smoothly as a function of $\chi_{1,z},\chi_{2,z}$ for
nonprecessing binaries of different mass ratios $q=m_1/m_2$.  
Targeted NR simulations have corroborated the simple dependence 
of the likelihood seen here.
Despite employing simulations with two strongly precessing spins and including higher harmonics,  two factors which have been
previously shown to be able to break degeneracies  \cite{2006PhRvD..74l2001L,2009PhRvD..80f4027K,2011PhRvD..84b2002L,gwastro-mergers-HeeSuk-FisherMatrixWithAmplitudeCorrections,gwastro-mergers-HeeSuk-CompareToPE-Precessing,gwastro-mergers-PNLock-Distinguish-Daniele2015}, Table
\ref{tab:SimulationRanks:30} reveals that simulations with the same values of $q$ and $\chi_{\rm eff}$ almost always have similar
values of $\ln L$.  In other words, these two simple parameters explain most of the variation in $L$, even when $L$
changes by up to a factor of $e^{100}$.   
Finally and critically,   simulations with similar physics produce very similar results.  By adopting $f_{\rm
  low}=30\unit{Hz}$ and thereby largely standardizing simulation duration, we find similar values of $\ln L$ when
comparing the data to simulations performed by different groups with similar (or even identical) parameters.

Our results and that of \PaperPE{}  constrain the progenitor binary's redshifted mass, mass ratio, and
aligned effective spin  $\chi_{\rm eff}$; see Table \ref{tab:ParameterRanges}.    The effective spin is defined as
   \cite{2001PhRvD..64l4013D,2008PhRvD..78d4021R}
\begin{eqnarray}
\chi_{\rm eff} = (\mathbf{S}_1/m_1  +\mathbf{S}_2/m_2)\cdot \hat{L}/M \;,
\end{eqnarray}  
 For example, the color scale in Figure \ref{fig:Constraints30:MQXi} provides
a graphical representation of $\ln L$ versus $\chi_{\rm eff}$; large values of $|\chi_{\rm eff}|$ (only possible for spin-aligned systems) are
inconsistent with the data. 
The agreement between our results and \PaperPE{}
 persists
despite using a much larger simulation set than those used to calibrate the models used in \PaperPE{}; and despite
employing simulations with black hole spins that are both precessing and with magnitude significantly outside the range
$\chi < 0.5-0.8$ for which these models were calibrated
\cite{gw-astro-EOBspin-Tarrachini2012,2015PhRvD..92j2001K,2015arXiv150807253K}.  

The three parameters $M_z$, $q$, and $\chi_{\rm eff}$
are well-known to have a strong and tightly-correlated impact on the gravitational wave signal
and hence on implied posterior distributions
\cite{1995PhRvD..52..848P,gw-astro-mergers-NRParameterEstimation-Nonspinning-Ajith,LIGO-CBC-S6-PE,gwastro-mergers-HeeSuk-CompareToPE-Aligned,HannamOhmeAlignedSpin}.
Since general relativity is scale-free, the total redshifted binary mass $M_z$ sets the characteristic physical
timescale of the coalescence.  Due to strong spin-orbit coupling, aligned spins ($\chi_{\rm eff} >0$) extend the temporal duration of the
inspiral \cite{ACST} and coalescence of the two black holes (e.g., the hangup effect
\cite{Campanelli:2006uy}); aligned spins also increase the final black hole spin and hence extend the duration of the post-merger
quasinormal ringdown \cite{2009CQGra..26p3001B}.  More extreme mass ratio extends the duration of the pre-merger phase while
dramatically diminishing the amplitude and frequency of post-merger oscillations
\cite{1998PhRvD..57.4535F,gwastro-nr-SemiAnalyticModel-McWilliams2010,nr-Jena-nonspinning-templates2007,gwastro-nr-Phenom-Lucia2010}.  
As noted above, 
the data tightly constrain one of these combinations (e.g., the total redshifted mass at fixed
simulation parameters).  
Hence, our ability to constrain any individual parameter $M_z,q$, or $\chi_{\rm eff}$ is limited not by the accuracy to which $M_z$
is determined for each simulation (i.e., the width $1/\sqrt{\Gamma_{MM}}$), but rather by differences between
simulations (i.e., trends in $\ln L$ versus $\chi_{\rm eff},q$) which break the degeneracy between these tightly correlated parameters.  
Simulations with a variety of physics fit the data, including strongly precessing systems.  In  Table
\ref{tab:SimulationRanks:30}, several simulations with large transverse spins but nearly zero net aligned spin  fit the data almost as well
as the best-fitting nonprecessing simulations (e.g., SXS:BBH:3; RIT simulation \verb|D10_q0.75_a-0.8_xi0_n100|).
As described below, Table \ref{tab:SimulationRanks:AllSignal} shows that these and other long precessing simulations fit even better when more
low-frequency content is included.
The  correspondence between our results and those presented in \PaperPE{} merits further reflection: by
construction our fiducial analysis (Table \ref{tab:SimulationRanks:30}) omitted nontrivial early-time information (i.e.,
$f<30\unit{Hz}$) which, for each simulation, more tightly constrains the
range of masses that could be consistent with the data.  
In fact, as we show below, strong degeneracies in the gravitational wave signal between mass, mass ratio, and spin imply
that our ability to break these degeneracies dominates our reconstruction of source parameters.  Omitting information
from low frequencies marginally reduces our ability  to identify  the range of masses that are consistent
with the data  \emph{for one simulation}; however, this omission does not impair our ability to draw conclusions overall, after accounting for uncertain spins and mass ratio.

Directly comparable to Fig.~12 in \PaperBurst{}, the right panel of Figure \ref{fig:Constraints30:MQXi}  provides a visual representation of one key correlation
between $q$ and $\chi_{\rm eff}$: only a narrow range of mass ratios and aligned
effective  spin $\chi_{\rm eff}$ are consistent with the data. 
This range includes both nonprecessing and precessing simulations.   
Most other parameters have a subdominant effect.  For example, restricting attention to nonprecessing binaries for clarity,
the data do not strongly
discriminate between systems with similar $\chi_{\rm eff}$ but different $\chi_{1,z},\chi_{2,z}$; see, e.g., Figure \ref{fig:ap:AlignedSpinMaxLmarg}.

\subsection{Results for nonprecessing sources, including interpolation}
\label{sec:sub:aligned}

\begin{figure*}
\includegraphics[width=0.33\textwidth]{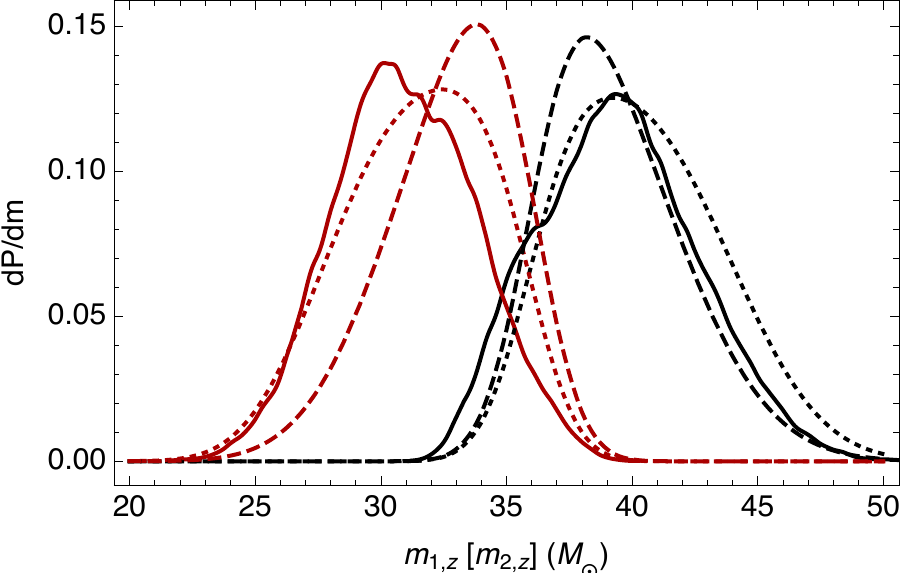}
\includegraphics[width=0.33\textwidth]{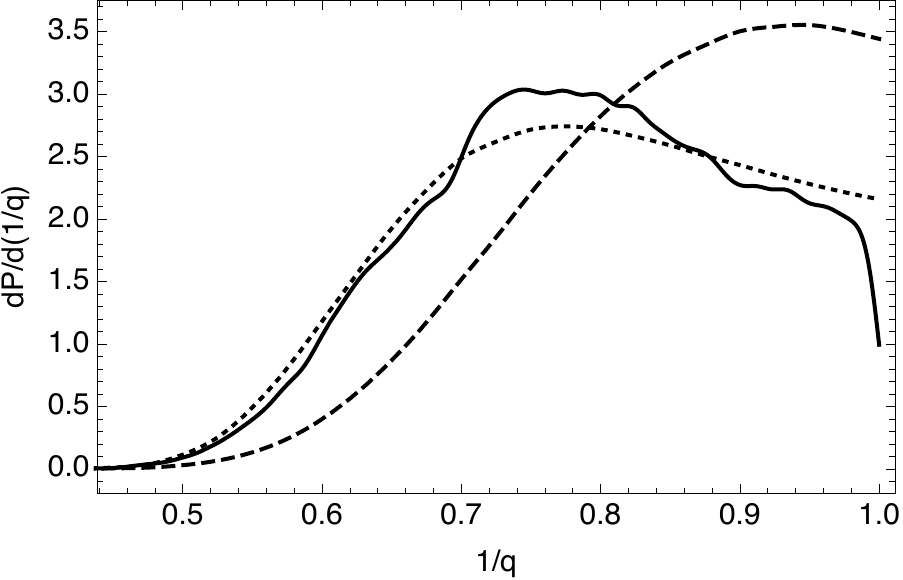}
\includegraphics[width=0.33\textwidth]{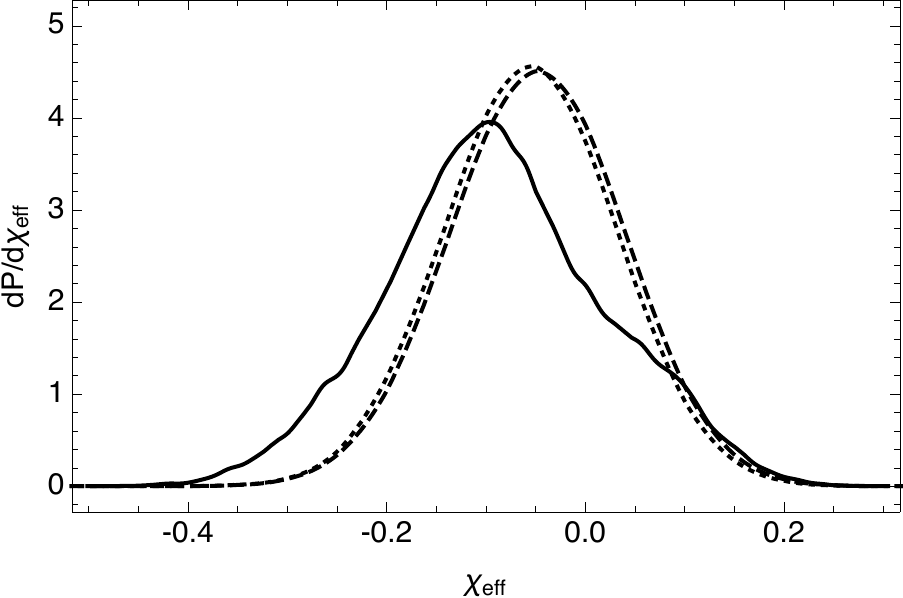}
\caption{\label{fig:PosteriorAgreement}\textbf{Distributions agree [nonprecessing case]}: Comparison between the
  posterior distributions reported in \PaperPE{} for nonprecessing binaries (solid) and the posterior distributions implied by
  a leading-order approximation to $\lnLmarg$    [Eq. (\ref{eq:maxL})] derived using $l\le 2$ (dotted) and $l\le 3$
   (dashed).
\emph{Left panel}:   $m_{1,z}$ (black) and $m_{2,z}$ (red).  \emph{Center panel}:  Mass ratio $1/q = m_{2,z}/m_{1,z}$.  The data increasingly favor comparable-mass binaries
as higher-order harmonics are included in the analysis.   
\emph{Right panel}:
Aligned effective spin $\chi_{\rm eff}$. 
The noticeable differences between our  $\chi_{\rm eff}$ distributions and the solid curve are also apparent in 
 Figures \ref{fig:Constraints30:Chi1zChi2z} and \ref{fig:Constraints30:MQXi}: 
  our analysis favors a slightly higher effective spin.
}
\end{figure*}

\begin{figure*}
\includegraphics[width=\textwidth]{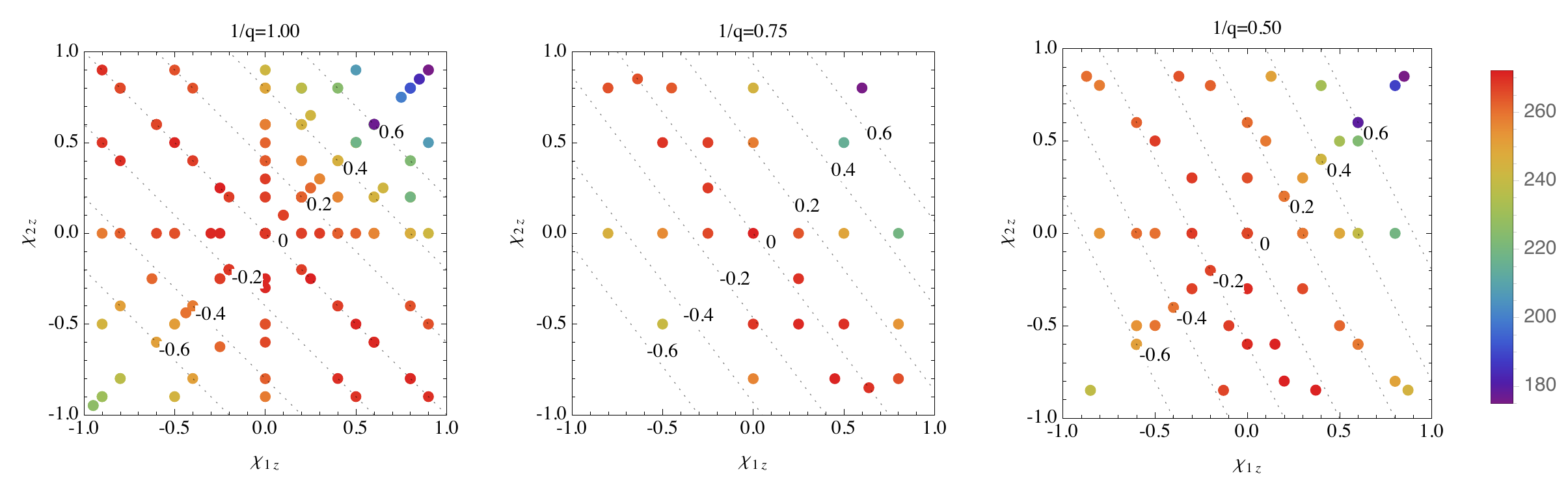}
\caption{\label{fig:ap:AlignedSpinMaxLmarg}\textbf{Likelihood versus  spins: Nonprecessing}: 
Maximum likelihood
  $\ln L$ (colors, according to the colorbar on the right)
  as a function of spins $\chi_{1,z},\chi_{2,z}$ 
  for different choices of mass ratio $1/q$, computed using
  all $l=2$ modes.   Each point represents a nonprecessing NR simulation
    from Table \ref{tab:SimulationRanks:30}.  To increase
  contrast,  simulations with $\ln L <
  171$ have been suppressed.  Only  simulations
  with $f_{\rm
    start}<30\unit{Hz}$ are included.    Dashed lines and labels indicate contours of constant $\chi_{\rm eff}$.    The
  left two panels show that for mass ratio $q\simeq 1$,
  the marginalized likelihood is approximately constant on lines 
  of constant $\chi_{\rm
    eff}$.  For more asymmetric binaries ($q=2$), the 
  marginalized likelihood is no longer 
  constant on lines of constant
  $\chi_{\rm eff}$.  Along lines of constant  $\chi_{\rm eff}$ and $q$, $\ln L$ decreases versus $\chi_{2,z}$
}
\end{figure*}

\begin{figure}
\includegraphics[width=\columnwidth]{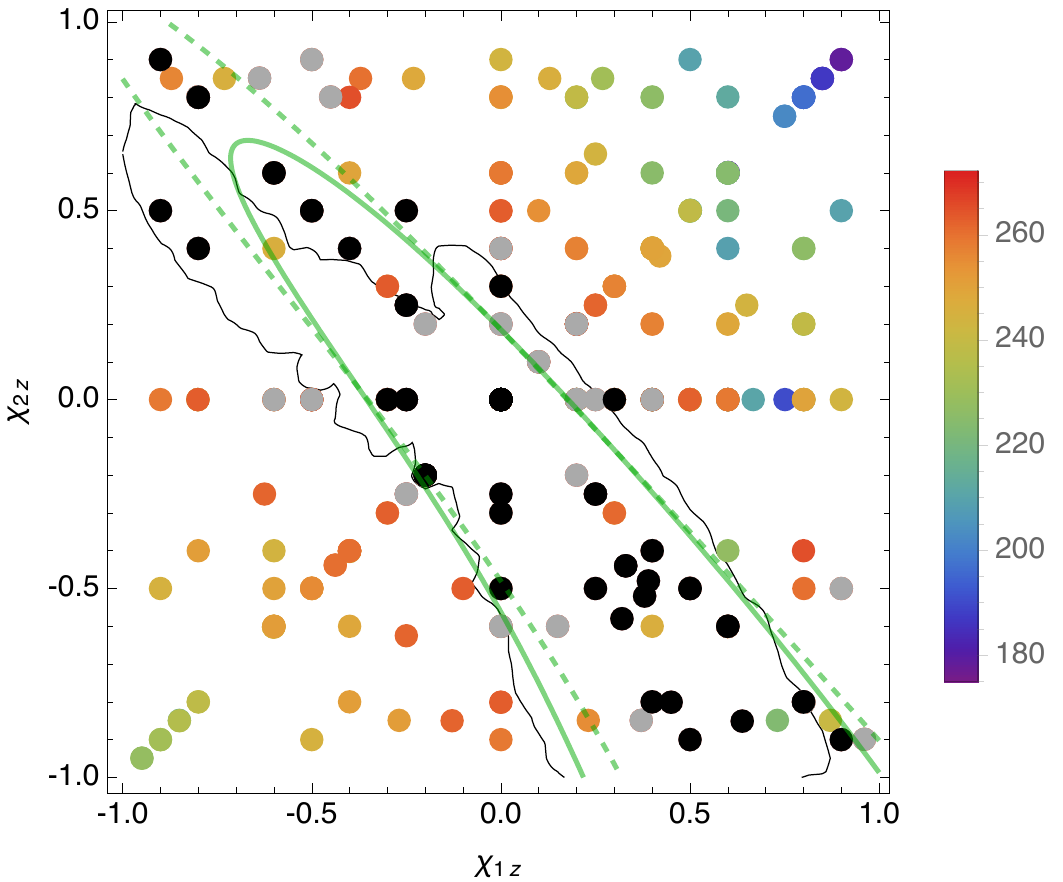}
\caption{\label{fig:Constraints30:Chi1zChi2z}\textbf{Aligned spin components not constrained [aligned only shown]}:  
  Colors represent the marginalized log likelihood as a function
  of the aligned spin components $\chi_{1,z}$ and $\chi_{2,z}$.
  Each point represents an NR simulation;  only  nonprecessing  simulations are included.
  Points with $\lnLmargGray < \ln L < \lnLmargBlack$ are shown in light gray,
  with $\ln L > \lnLmargBlack$ are shown in black, and 
  with $\ln L < \lnLmargGray$ are shown according to the
  color scale on the right (points with $\lnLmarg< 172$
  have been suppressed to increase contrast).   [The quantity $\ln L$ is the maximum value of $\lnLmarg$ with respect to
    mass; see Eq.~(\ref{eq:lnL}).]
  Consistent with our other results,  $f_{\rm low}=30\unit{Hz}$.
  For
  comparison, the solid \TheColorSEOB{} contours show the \pCI{}
  credible intervals derived in \PaperPE{}, assuming spin-orbit alignment and omitting corrections for waveform
  systematics.     The solid and dashed green contours are the nominal \pCI{} credible interval derived using an approximation to
  our data for $\lnLmarg$, assuming both spins are exactly parallel to the orbital angular
  momentum, for $l=2$ (solid) and $l=3$ (dashed), respectively; see Section \ref{sec:sub:aligned} for more details.
}
\end{figure}

Both \PaperPE{} and our highly-ranked simulations
in Table \ref{tab:SimulationRanks:30} demonstrate that binary black holes
with nonprecessing spins can reproduce \TheEvent{}.
Only four
parameters characterize a nonprecessing binary:
the two component masses $m_1,m_2$ and the components of each BH's dimensionless spin $\boldsymbol{\chi}_i$ projected perpendicular to the
orbital plane ($\chi_{1,z},\chi_{2,z}$).   Nonprecessing
binary black hole coalescences have been extensively
simulated \cite{2010RvMP...82.3069C}; see, e.g., Table \ref{tab:SimulationList}.  Several  models have been developed to reproduce the leading-order gravitational wave
emission (the $(l,|m|)=(2,2)$ modes)
\cite{2014PhRvD..89f1502T,2014PhRvD..89h4006P,nr-Jena-nonspinning-templates2007,gwastro-nr-Phenom-Lucia2010,gwastro-mergers-IMRPhenomP,2015PhRvL.115l1102B}; 
one, the SEOBNRv2 model \cite{gw-astro-EOBspin-Tarrachini2012}, is adopted as a
fiducial reference by
\PaperPE.   While this model 
has not been calibrated to NR for large values of $\chi_{\rm eff}$ and $q$ \cite{2015PhRvD..92j2001K}, it
has been shown to accurately reproduce the $(2,2)$ mode from binaries with comparable
mass and low spins \cite{2013CQGra..31b5012H,2015PhRvD..92j2001K,gwastro-mergers-nr-PrayushAlignedSpinAccuracy-2016}.  
  Because of degeneracies, data from \TheEvent{} do not easily distinguish
  between different points in parameter space that have the same values
  of $M_z,q,\chi_{\rm eff}$; in particular, it is difficult to individually
  measure $\chi_{1,z}$ and $\chi_{2,z}$ when $q\simeq 1, \chi_{\rm eff} \simeq 0$ and $\chi_{1,z}\simeq - \chi_{2,z}$; see, e.g.,  \cite{HannamOhmeAlignedSpin}.
  Because \TheEvent{} has comparable masses and is oriented face-off with
  respect to the line of sight, even including higher-order modes in the
  gravitational waveform (which we do in our approach here but is not done
  for the analytic waveform models)
  does not strongly break these degeneracies and allow
  us to distinguish individual spins.

 By 
 stitching together our fits for $\lnLmarg(M_z)$ and reconstructing the relevant parts of the likelihood
for all masses and aligned spins, we can estimate the full posterior distribution for $M_z,q,\chi_{1,z},\chi_{2,z}$  using Eq. (\ref{eq:Posterior}).  Due to inevitable systematic modeling errors in the fit, as described below, this approximation may  not
have the statistical purity of the method presented in \PaperPE{}: any credible intervals or deductions drawn from it
should be interpreted with judicious skepticism.   On the other hand, this method enables the reader to recalculate the
posterior distribution using any prior $p(\lambda)$, including astrophysically-motivated choices. 
Fitting to nonprecessing simulations, we find  $\lnLmarg$  for $\lnLmarg>262$ is reasonably well-approximated by a quadratic function of
the intrinsic parameters $\mc_z = (m_{1,z} m_{2,z})^{3/5}/(m_{1,z}+m_{2,z})^{1/5}$, $\eta = (m_{1,z} m_{2,z})/(m_{1,z} +
m_{2,z})^{2}$, $\delta = (m_{1,z}-m_{2,z})/M_z$, $\chi_{\rm eff}$, and $\chi_- \equiv (m_{1,z}\boldsymbol{\chi}_1-m_{2,z} \boldsymbol{\chi}_2)\cdot \hat{\mathbf{L}}/M_z$:
\begin{subequations}\label{eq:maxL}
\begin{eqnarray}
\lnLmarg \simeq 268.4  - \frac{1}{2}(\lambda-\lambda_*)_a\Gamma_{ab}(\lambda-\lambda_*)_b - \Gamma_{\chi_{-}\delta} \chi_{-}\delta.
\end{eqnarray}
where the indices $a,b$ run over the variables 
$\mc_z,\eta,\chi_{\rm eff},\chi_-$.
In this expression, $\lambda_a$ represents the vector 
$(\mc_z,\eta,\chi_{\rm eff},\chi_-)$,
 $\lambda_{*a}$ corresponds to the vector
($\mc_z=31.76 M_\odot$, $\eta=0.255$,  $\chi_{\rm eff}=-0.037$, $\chi_-=0$) of parameters which maximize $\lnLmarg$, and
$\Gamma$ 
is a matrix (indexed by $\mc_z,\eta,\chi_{\rm eff},\chi_-,\delta$) with
numerical values
\begin{eqnarray}
\Gamma = \begin{bmatrix}
 3.75 & -224.2 & -52.0 & 0  & 0 \\
 -224.2 & 22697.2 & 2692 & 0 & 0\\
 -52.0 & 2692. & 846.9 & 0  & 0\\
 0 & 0 & 0 & 2.57  & -16.3 \\
 0 & 0 & 0 & -16.3  &  0 \\
\end{bmatrix}.
\end{eqnarray}
\end{subequations}
Here
we retain many significant digits to account for structure in $\Gamma$, which is
nearly singular.  
 Equation (\ref{eq:maxL})
respects exchange symmetry $m_{1,z},\chi_1 \leftrightarrow m_{2,z},\chi_2$.
Our results do not sensitively depend on the value of $\Gamma_{\chi_-,\chi_-}$,
indicating that this quantity is not strongly constrained by the data.
Conversely, the posteriors do depend 
on $\Gamma_{\chi_-,\delta}$.  
As the contrast between the first term in
Eq. (\ref{eq:maxL}) and the data Table \ref{tab:SimulationRanks:30} makes immediately apparent, this  coarse
approximation  can differ from the simulated results by of order $1.7$ in the log (rms residual).  This reflects
the combined impact of  Monte Carlo error, systematic   error
caused by too few orbits in some simulations,  and systematic errors caused by sparse placement of
NR simulations and non-quadratic behavior  of 
$\lnLmarg$ with respect to parameters.  
Repeating our calculation while including all the $l \le 3$ modes, we find 
the same functional form as Eq.~(\ref{eq:maxL}), but with a different vector
$\lambda_{*,a}^{(3)}$=
($\mc_z=38.1 M_\odot$, $\eta\simeq 0.32$,  $\chi_{\rm eff}=0.11$, $\chi_-=0$), 
and  a different matrix
\begin{eqnarray}
\Gamma^{(3)} = \begin{bmatrix}
 3.746 & -235.5 & -51.5 & 0  &0 \\
 -235.5 & 17970 & 2941 & 0 & 0  \\
 -51.5 & 2941 & 833.2 & 0 & 0 \\
 0 & 0 & 0 & 0.57 & -12.57\\
 0 & 0 & 0 & -12.57 &  0\\
\end{bmatrix}.
\end{eqnarray}
We label $\lambda^{(3)}$ and $\Gamma^{(3)}$ with a  superscript ``3''  to distinguish this result from the
corresponding result using only $l=2$ modes shown in Eq.~(\ref{eq:maxL}).

For nonprecessing sources, using Eq. (\ref{eq:Posterior}) and a uniform prior in $\chi_{1,z},\chi_{2,z}$ and the two component masses, we can
evaluate the marginal posterior probability $p(z)$ for any intrinsic parameter(s) $z$.    The two-dimensional marginal posterior probability is
shown as a green solid ($l=2$) and dashed ($l \le 3$) line in Figures
\ref{fig:Constraints30:MQXi} and 
\ref{fig:Constraints30:Chi1zChi2z}.
Both the $l=2$ and $l\le 3$ two-dimensional
distributions are in reasonable agreement with the posterior distributions reported in \PaperPE{} for nonprecessing
binaries, shown  as a \TheColorSEOB{} curve in these
figures.  
These two-dimensional distributions are also consistent with the 
distribution of simulations with $\lnLmarg > \lnLmargBlack$  (i.e., black
points).  
Additionally,  Figure \ref{fig:PosteriorAgreement} shows several one-dimensional marginal probability distributions
($m_{1,z},m_{2,z},q,\chi_{\rm eff}$),
shown as   dotted   ($l=2$) or dashed lines
($l\le 3$); for comparison, the solid line shows the corresponding
distribution from \PaperPE{} for nonprecessing binaries.

Despite broad qualitative agreement, these comparisons highlight several differences between our results and \PaperPE{},
and between results including $l=2$ modes and those including all $l\le 3$ modes.  
For example, Figure \ref{fig:Constraints30:MQXi} shows that
the distribution in  $M_z,q,\chi_{\rm eff}$, computed using our
method (solid green lines and black points) is slightly different than the corresponding distributions in \PaperPE{}.
As seen in this figure and in Figure \ref{fig:Constraints30:Chi1zChi2z}, 
the posterior distribution in \PaperPE{} includes
binaries with low effective spin, outside the support of the distributions reported here.   These differences are
directly reflected in the marginal posterior  $p(\chi_{\rm eff})$  (right panel of  Figure
  \ref{fig:PosteriorAgreement}) and in Table \ref{tab:ParameterRanges}.  
Our results for the component spins $\chi_{1,z},\chi_{2,z}$, the effective spin $\chi_{\rm eff}$,  the total mass
$M_z$, and the mass of the more massive object $m_{1,z}$ do not change significantly when $l=3$ modes are included.  
The mass  ratio distribution $p(q)$ is
also slightly different from \PaperPE{} when $l=3$ modes are included;  see Figure  \ref{fig:PosteriorAgreement}.  Compared to prior work, this analysis favors
comparable-mass binaries when higher modes are included; see, e.g., the center panel of  Figure  \ref{fig:PosteriorAgreement}.

The differences between the results reported here and \PaperPE{} should be considered in context: not only does our
study employ  numerical relativity without analytic waveform models, 
but it also adopts a slightly different starting frequency, omits any direct
treatment of calibration uncertainty, and employs a quadratic approximation to the likelihood. 
That said, comparisons conducted under similar limitations and using real data, differing only in the underlying waveform model, reproduce
results from \textsc{LALInference};  see \PaperPENRMethods{} for details.
By assuming the binaries are strictly aligned but permitting generic spin 
magnitudes, our analysis (and that in \PaperPE{}) neglects prior information
that  could be used to significantly influence the posterior spin
distributions.
For example,  the part of the posterior in the bottom
right quadrant of Figure
\ref{fig:Constraints30:Chi1zChi2z} is unstable to large angle precession
\cite{gwastro-mergers-PNLock-UpDownInstability}: if a comparable-mass binary formed at large separation with
$\chi_{1,z}>0$ and $\chi_{2,z}<0$, it could not remain aligned 
during the last few orbits.
Likewise, the astrophysical scenarios most likely to produce strictly
aligned binaries --- isolated binary evolution --- are most likely 
to  result in both $\chi_{1,z},\chi_{2,z}>0$: both
spins would be strictly and positively 
aligned (see, e.g, \cite{2016arXiv160204531B}). %
In that case, only the top right quadrant of Figure
\ref{fig:Constraints30:Chi1zChi2z} would be relevant.
Using the analytic tools provided here, the reader can regenerate approximate posterior distributions employing any
prior assumptions, including these two   considerations.

\subsection{Transverse and precessing spins}

\begin{figure*}
\includegraphics[width=\columnwidth]{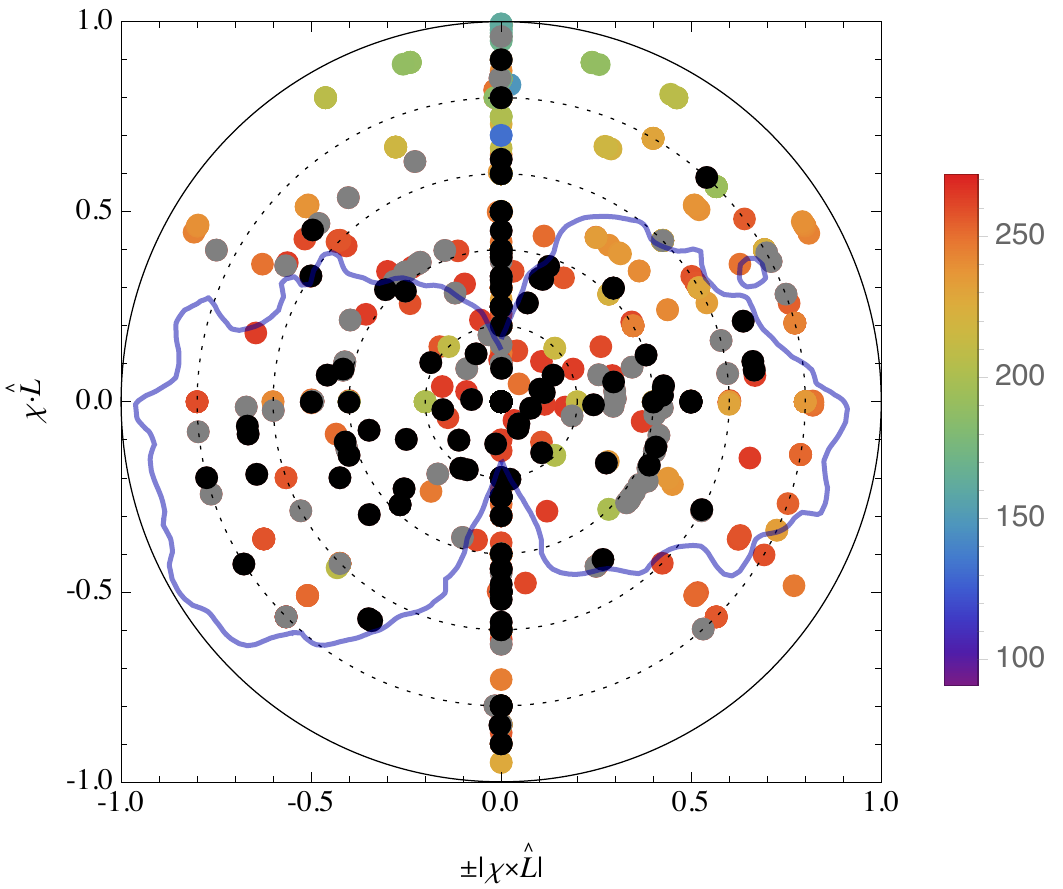}
\includegraphics[width=\columnwidth]{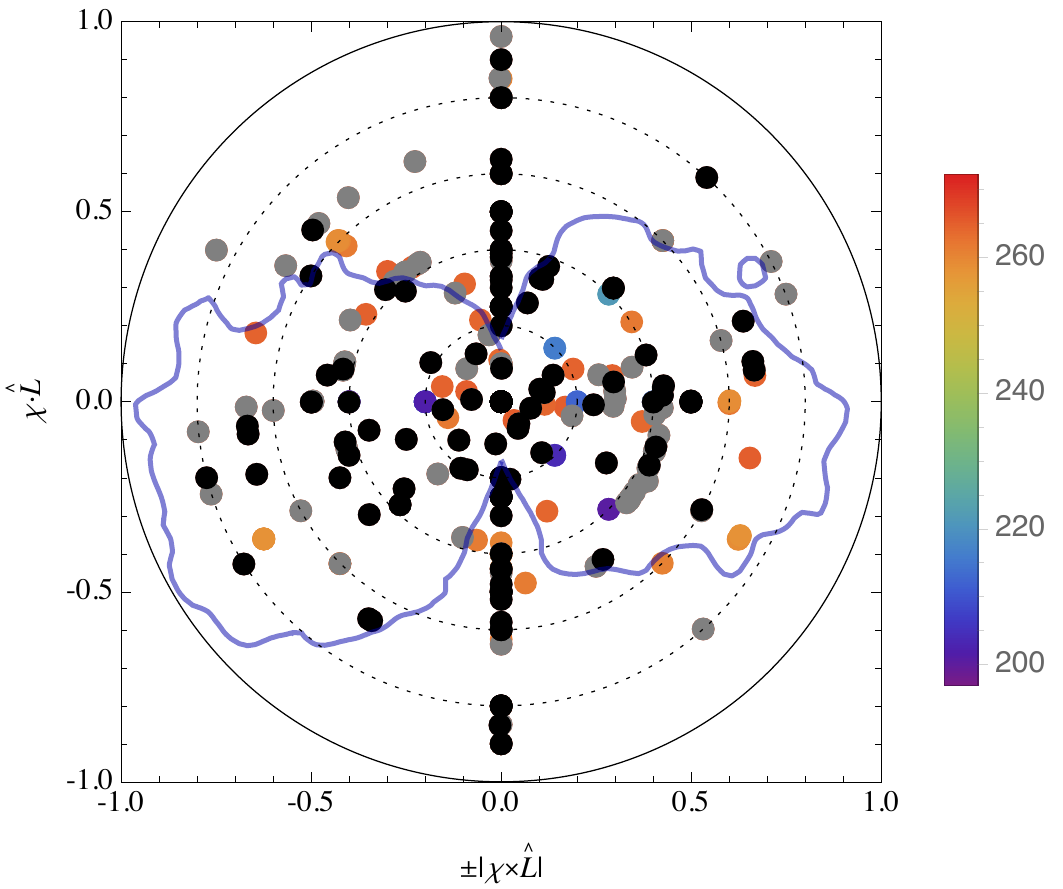}
\caption{\label{fig:SpinDiskPlot}\textbf{Large spins possible}:  
  Colors represent marginalized log likelihood as a function
  of
  $(-1)^{i+1}|\boldsymbol{\chi}_{i}\times \hat{\mathbf{L}}|$ 
   and
  $\boldsymbol{\chi}_{i}\cdot \hat{\mathbf{L}}$, where
  $i=1,2$ indexing the first and second black hole,evaluated using each simulation's initial conditions [Table
    \ref{tab:SimulationList}]; compare to the left panel of Figure 5 in \PaperPE.   Each simulation
   therefore appears twice in this figure: once on the left and once on the right. 
Spin magnitudes and directions refer
  to the initial configuration of each NR simulation, not to properties at a fixed reference frequency as in
  \PaperPE{}.   
  Points with 
    $\lnLmargGray < \ln L < \lnLmargBlack$
  (cf Eq. (\ref{eq:DefineGrayBlackCutoffs})) are shown in light gray,
  with $\lnLmarg > \lnLmargBlack$ are shown in black, and 
  with $\lnLmarg < \lnLmargGray$ are shown according to the
  color scale on the right.
[The quantity $\ln L$ is the maximum value of $\lnLmarg$ with respect to
    mass; see Eq.~(\ref{eq:lnL}).]
  While this figure
  was evaluated using $l=2$ modes only, the corresponding figure for $l\le 3$ modes is effectively indistinguishable. 
  This diagram demonstrates that both black holes could have large dimensionless spin $\chi$.  The solid black circle represents the
  Kerr limit $|\boldsymbol{\chi}|=1$; to guide the eye, the dashed circles show $|\boldsymbol{\chi}_{1,2}| = 0.2, 0.4,
  0.6, 0.8$.
   For comparison, the \TheColorIMRP{} contour shows the corresponding $\pCI$ credible
  interval  reported in \PaperPE, using spin configurations at 20\unit{Hz}.
     The structure in this contour (e.g., the absence of support near the axis) should not be over-interpreted: similar structure
  arises when reconstructing the parameters of synthetic nonprecessing sources.
 \emph{Left panel}: All simulations are included.
\emph{Right panel}: 
To increase contrast
only simulations with  $q<2$ and $\chi_{\rm eff} \in[-0.5,0.2]$ are shown; these limits are chosen to be
  consistent with the two-dimensional posterior 
   in $q,\chi_{\rm eff}$ shown
   in the right panel of Figure~\ref{fig:Constraints30:MQXi}.
}
\end{figure*}

Figure \ref{fig:SpinDiskPlot} shows the maximum likelihood for
the available NR simulations, plotted as a function of the magnitude
of the aligned and transverse spin components.  The figure shows that
there are both precessing and nonprecessing simulations that have
large likelihoods (black points), indicating that many precessing
simulations are as consistent with the data as nonprecessing
simulations.  Moreover, simulations with large precessing spins are
consistent with the \TheEvent{}: many configurations have $\chi_{\rm eff}
\simeq 0$ but large spins on one or both BHs in the binary.
Keeping in mind the limited range
of simulations available, the magnitude and direction of either BHs spin cannot be significantly constrained by our method.

\begin{figure}
\includegraphics[width=\columnwidth]{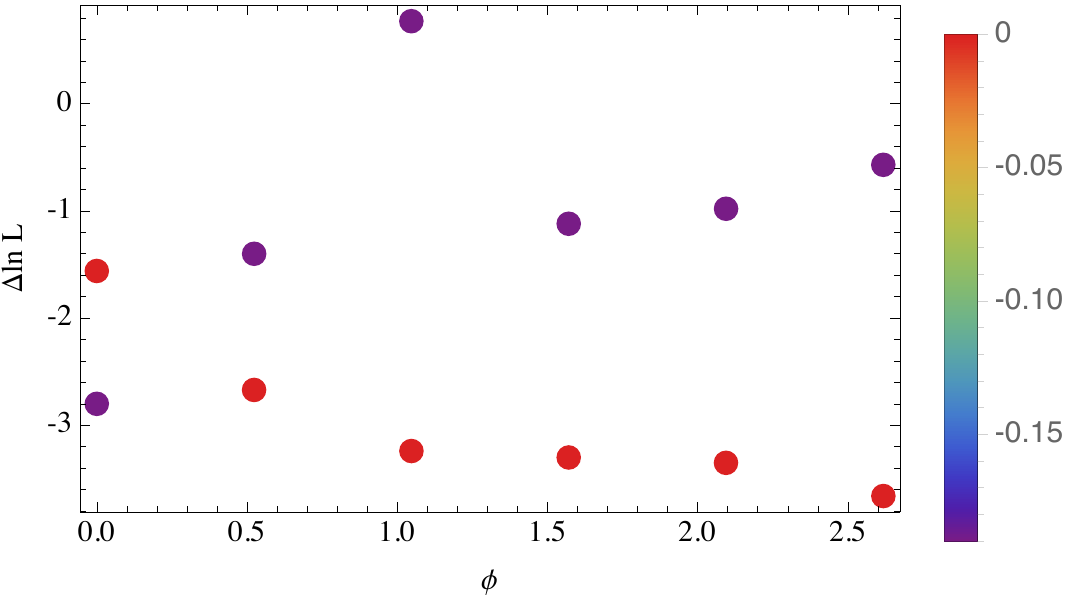}
\caption{\label{fig:Spin:TransverseDependencePossible}\textbf{Transverse spins can influence the marginal likelihood}:
$\Delta \ln L$, the difference between the computed $\ln L$ of a
        precessing simulation (see Eq.~(\ref{eq:lnL})) and the estimated
        value of $\ln L$ from our fit to nonprecessing simulations, plotted
        as a function of an angle $\phi$, for 
        two one-parameter families of simulations whose initial 
        conditions differ only by a rotation of the initial spins
        through an angle $\phi$ around the
        initial angular momentum axis.
The color scale indicates 
  the value of $\chi_{\rm eff}$.
}
\end{figure}

Not all precessing simulations with suitable $q,\chi_{\rm eff}$
are consistent with \TheEvent{}; some have values of $\ln L$ 
that are not within 10 of the peak; 
see the right panel of Fig.~\ref{fig:SpinDiskPlot})
The marginal
log-likelihood $\ln L$ depends on the transverse spins, 
not just the dominant parameters ($q,\chi_{\rm eff},\mc_z$).    
As a concrete illustration, Figure \ref{fig:Spin:TransverseDependencePossible} shows that the marginalized log likelihood depends on the specific direction of the transverse spin, in the plane
perpendicular to the angular momentum axis.    Specifically, this figure compares the peak marginalized log likelihood
($\ln L$) calculated for each simulation with the value of $\ln L$
predicted from our fit to nonprecessing binaries.  For
precessing binaries,  $\ln L$
 is neither in perfect agreement with the nonprecessing prediction, 
nor independent
of  rotations of the initial spins about the initial orbital
angular momentum by an angle  $\phi$.
While the transverse spins do influence the likelihood, slightly, 
 the data do not
favor any particular precessing configurations.    
No precessing simulations had
marginalized likelihoods that were both significant overall and 
significantly above the value we predicted assuming
aligned spins.  In other words, the data do 
not seem to favor precessing systems, when analyzed using only information
above 30 Hz. 

Our inability to  determine
the most likely transverse spin components
is expected, given both our self-imposed restrictions ($f_{\rm low}=30\unit{Hz}$) and the a
priori effects of geometry.    
For example, the
lack of apparent modulation in the 
signal reported in \PaperDetection{} and  \PaperBurst{} points to an orientation with $\mathbf{J}$
parallel to the line of sight, along which precession-induced modulations are highly suppressed.    In addition,
the high mass and hence extremely short observationally accessible signal above $10\unit{Hz}$ provides relatively few
cycles with which to extract this information.  The timescales involved are particularly unfavorable to attempts to
extract precession-induced modulation from the pre-merger signal: the pre-coalescence precession rate for these sources
is low ($\Omega_p\simeq (2 + 3 m_2/m_1) J/2 r^3 \simeq 2\pi\times 1\unit{Hz}(f/40\unit{Hz})^{5/3}$ for this system,
where $J$ is the magnitude of the total orbital angular momentum and we assume $J\simeq L$; see \cite{ACST}), implying at best two pre-merger precession cycles could be accessible
from the early signal; see \PaperPE.   As with the total binary mass, spin
information will be accessible at lower
frequencies (i.e., between $10-30\unit{Hz}$); however,
our fiducial analysis using $f_{\rm low}=30\unit{Hz}$ is not well-suited to extract it.
For a suitably-oriented source, the strongly nonlinear merger phase  can in principle encode significant information about the coalescing binary's precessing
spins.  Qualitatively speaking, this information is encoded in the relative amplitude and phase of subdominant
quasilinear perturbations, causing the radiation from the final black hole to appear to precess
\cite{gwastro-mergers-nr-Alignment-ROS-CorotatingWaveforms,gwastro-mergers-nr-Alignment-ROS-Polarization}.  
This information also influences the final black hole mass and spin.    The model used in \PaperPE{} adopted a geometric
ansatz to incorporate these effects at leading order, using a lower-dimensional effective model for a single precessing spin.  
However, in this work, despite including higher modes and having direct access to  as-yet unmodeled effects, our analysis shows no significant
difference from the previously reported conclusions regarding the transverse spin distribution.
The low frequency content of GW150914 may contain  some further signature consistent
with two precessing spins.   Simultaneously with this work, an analysis has been performed using semianalytic models that can fully capture
both spins' dynamics  \PaperSEOBNRvthree{}. %
Within the context of this study,  Table \ref{tab:SimulationRanks:AllSignal} shows an analysis without  an artificially imposed low frequency cutoff.  As expected, the best-fitting long simulations seen in our previous report
fit equally well and agree.  Notably, however, we find an increase in the marginalized likelihood for
precessing simulations like  \verb|SXS:BBH:308| 
and \verb|D21.5_q1_a0.2_0.8_th104.4775_n100|.    More broadly, when we
include low-frequency content, many precessing simulations that previously had not fit the high-frequency content as
well  become  more significant.  
However, to extract low-frequency content reliably, we will need to both hybridize these precessing simulations and
interpolate the likelihood as a function of both precessing spins.  These further investigations are beyond the scope of
the present study.

\section{Results II: Strong-field properties and post-coalescence parameters}
\label{sec:Results:FinalState}

\begin{figure}
\includegraphics[width=\columnwidth]{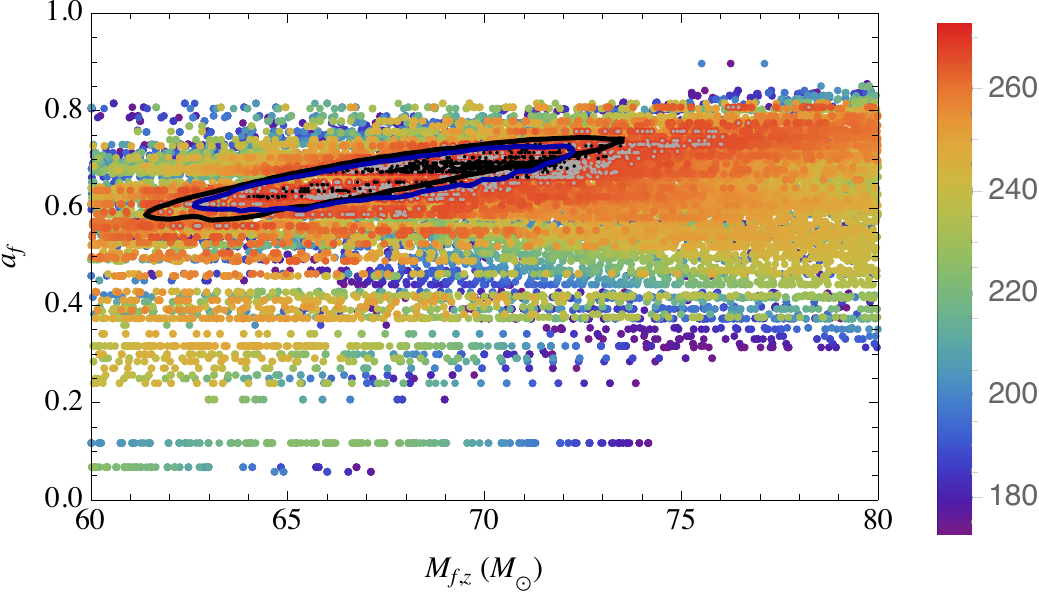}
\caption{\label{fig:FinalState}\textbf{Final redshifted mass and spin}: 
  The final redshifted black hole masses $M_{f,z}$ and
  spins $a_f$. 
   Each point represents an NR simulation; both nonprecessing and precessing simulations are included.
  Points with $\lnLmargGray < \lnLmarg < \lnLmargBlack$ are shown in light gray,
  with $\lnLmarg > \lnLmargBlack$ are shown in black, and 
  with $\lnLmarg < \lnLmargGray$ are shown according to the
  color scale on the right (points with $\lnLmarg < 172$
  have been suppressed to increase contrast).  
For comparison, the solid \TheColorSEOB{} curve shows the \pCI{} credible interval on $M_{f,z}$ and $a_f$ derived in \PaperPE{} and
\PaperTestGR{} using a spin-aligned model; the \TheColorIMRP{} curve shows the corresponding result derived from a
single-spin precessing (IMRP) model.  
}
\end{figure}

The numerical relativity simulations listed in Table \ref{tab:SimulationList} have been previously used to develop
accurate models for the final black hole mass and spin  \cite{Healy:2014yta,2008PhRvD..78d4002R,2011PhRvD..84l4052P,2015arXiv150807250H}.  The relationships developed in \cite{Healy:2014yta} for nonprecessing binaries were used
in  \PaperPE{} and \PaperTestGR{}  to infer the final black hole mass and spin, based on the pre-coalescence spins.  
By construction, this approximation neglects the impact of transverse spins.  Both  this work and in \PaperPE{} have
shown 
that \TheEvent{} is consistent both with nonprecessing and precessing pre-coalescence
spins.  When large, these spins are well-known to significantly
impact the final black hole mass and spin  \cite{2008PhRvD..77b6004B,gr-nr-io-fitting-Boyle2007,2010CQGra..27k4006L,2012ApJ...758...63B,2014PhRvD..89j4052L,2015PhRvD..92b4022Z}.

With direct access to both an accurate multimodal waveform for generic precessing systems and the final black hole
state, the method applied in this work is uniquely well equipped to identify the final black hole mass and spin.  
Figure \ref{fig:FinalState} shows our results.  Rather than approximate  a posterior distribution --- a
significant challenge in 8 dimensions --- we simply report  sets of points $M_{f,z},a_f$ corresponding to simulations
and initial redshifted masses $M_z$ so $\lnLmarg(M_{z})$ is greater than 
some cutoff.  When we include only
nonprecessing simulations, we find results consistent with the reported values in \PaperPE{} and \PaperTestGR{}.  
While many simulations listed in Table \ref{tab:SimulationRanks:30} have
 some transverse spin, many also have zero transverse spin, so overall the transverse spin distribution of our simulations is more concentrated
towards zero than the prior adopted in \PaperPE.  
Given the excellent agreement between our results and \PaperPE{} for pre-coalescence parameters, particularly in the
subset of spin-aligned binaries, we 
cannot identify any nonzero difference for final parameters that is
introduced by our methodology (e.g. our restriction to
$f_{\rm low}=30\unit{Hz}$).   

\section{\label{sec:Conclude}Conclusions}
Using a full Bayesian parameter estimation technique, we directly compare \TheEvent{} with a large set of binary black
hole simulations produced using full numerical relativity.   
Our comparisons employ physics and radiation content  ($l \le 3$ modes) not
available or only partially captured by the two semianalytic models used in \PaperPE{}.    
Using our completely independent approach, we nonetheless arrive at results similar to those of \PaperPE{}.  
Comparisons  including only the dominant modes (all $l\le 2$)    constrain the total redshifted mass $M_z$
[$64-82M_\odot$], mass ratio $1/q=m_2/m_q \in[0.6,1]$,
and effective aligned spin $\chi_{\rm eff} \in [-0.3, 0.2]$.    Including $l=3$ modes, we find the 
mass ratio is even more tightly constrained.
Both nonprecessing and precessing simulations fit the data; no compelling evidence exists for or against a
precessing origin.     Even accounting for precession, simulations with extreme mass ratios and
effective spins are highly inconsistent with the data, at any mass.  
Several nonprecessing and precessing simulations with similar mass ratio and $\chi_{\rm eff}$ are consistent with the
data.
Though correlated, the component spins (both in magnitude and direction) are not significantly constrained
  by the data:  the data are consistent with simulations with component spin magnitudes
   $a_{1,2}$ up to at least $0.8$, with random orientations.

This paper also provides the first concrete illustration, using real gravitational wave data,  of several methods to aid the interpretation
of gravitational wave  observations using numerical relativity.   
First and foremost, this method demonstrates that the marginalized likelihood can be efficiently evaluated on a grid
\cite{gwastro-PE-AlternativeArchitectures,2015CQGra..32w5017H}.   Straightforward reconstructions (e.g., fits, interpolation)
allow us to reconstruct the posterior at low cost.  
Further,  NR simulations are sufficiently dense, and the marginal log-likelihood $\lnLmarg$ sufficiently simple, that  $\lnLmarg$
 can be effectively approximated using  available
catalogs of NR simulations.  
Second, we provide and employ a simple but effective approximation to the marginalized likelihood.
A particularly efficient way to communicate results, this data product enables further investigations, including the
impact of the prior on our conclusions; the ability to incorporate the spin-precession instability into our
posterior \cite{gwastro-mergers-PNLock-UpDownInstability}; and anything involving conditional distributions, which are
trivially produced using the fit.  
Third, this investigation has  demonstrated the critical role that  numerical relativity can play in data analysis
while simultaneously illuminating a path forward in the era of frequent detections.  
We demonstrate that NR results can be directly applied to data analysis, without intervening
approximations.    In the future, while low-frequency sensitivity will improve, so will our ability to effectively hybridize
these simulations, so this approach will remain valuable even when very long signal models are required to reproduce the
data. 
Targeted followup can be performed guided by $\ln L$, our  measure of overall fit (maximizing $\lnLmarg$ over mass). 
Fourth, as described in \PaperPENRMethods{}, this method provides a direct and unambiguous method to assess the relative impact of
higher harmonics, waveform extraction, and modeling uncertainty on a point-by-point basis.    Investigations using this
technique will provide a valuable complement to parallel studies with \textsc{LALInference} \cite{LIGO-Puerrer-NR-LI-Systematics}.

As noted in \PaperAstro, the inferred spin magnitudes and misalignments provide unique and distinctive clues to the
astrophysical origin of \TheEvent{}.  
Notably, strongly misaligned spins require a violent origin, either through exceptionally dynamic stellar processes or a cluster
origin.  Our analysis cannot definitively support or rule out such an origin.   
 We recommend further analysis of \TheEvent{} with improved models for binary inspiral and
coalescence, whether derived semianalytically or via hybridization and/or interpolation of pure numerical relativity.  
For example, \PaperSEOBNRvthree{} reports marginally tighter constraints on (two) precessing spins, by comparing
\TheEvent{} against a model for the emitted radiation including  the very early inspiral, which by necessity NR simulations must omit.
Combined with this method, we further anticipate a large-scale simulation campaign in full numerical relativity to explore simulations comparable
to  \TheEvent{} could allow us to extract more insight into its nature. 
\PaperPENRMethods{} will provide further details on and examples with the method employed in this work.

\begin{acknowledgements}
The authors gratefully acknowledge the support of the United States
National Science Foundation (NSF) for the construction and operation of the
LIGO Laboratory and Advanced LIGO as well as the Science and Technology Facilities Council (STFC) of the
United Kingdom, the Max-Planck-Society (MPS), and the State of
Niedersachsen/Germany for support of the construction of Advanced LIGO 
and construction and operation of the GEO600 detector. 
Additional support for Advanced LIGO was provided by the Australian Research Council.
The authors gratefully acknowledge the Italian Istituto Nazionale di Fisica Nucleare (INFN),  
the French Centre National de la Recherche Scientifique (CNRS) and
the Foundation for Fundamental Research on Matter supported by the Netherlands Organisation for Scientific Research, 
for the construction and operation of the Virgo detector
and the creation and support  of the EGO consortium. 
The authors also gratefully acknowledge research support from these agencies as well as by 
the Council of Scientific and Industrial Research of India, 
Department of Science and Technology, India,
Science \& Engineering Research Board (SERB), India,
Ministry of Human Resource Development, India,
the Spanish Ministerio de Econom\'ia y Competitividad,
the Conselleria d'Economia i Competitivitat and Conselleria d'Educaci\'o, Cultura i Universitats of the Govern de les Illes Balears,
the National Science Centre of Poland,
the European Commission,
the Royal Society, 
the Scottish Funding Council, 
the Scottish Universities Physics Alliance, 
the Hungarian Scientific Research Fund (OTKA),
the Lyon Institute of Origins (LIO),
the National Research Foundation of Korea,
Industry Canada and the Province of Ontario through the Ministry of Economic Development and Innovation, 
the National Science and Engineering Research Council Canada,
the Brazilian Ministry of Science, Technology, and Innovation,
the Leverhulme Trust, 
the Research Corporation, 
Ministry of Science and Technology (MOST), Taiwan
and
the Kavli Foundation.
The authors gratefully acknowledge the support of the NSF, STFC, MPS, INFN, CNRS, and the
State of Niedersachsen/Germany  for provision of computational resources.

The SXS collaboration also gratefully acknowledges Compute Canada,  the Research Corporation, and California State
University Fullerton for computational
resources, as well as the the support of the National Science Foundation, the Research Corporation for Science
Advancement,and the Sherman Fairchild
Foundation.

The RIT team gratefully acknowledges the NSF for financial support, as well as Blue Waters and XSEDE for computational resources.

This paper has been assigned the document number LIGO-P1500263.
\end{acknowledgements}

\bibliography{references,paperexport,GW150914_refs}

\appendix

\clearpage
\begin{widetext}

\section{Simulation list}
\label{ap:FullSimulationDetails}
In this section, we enumerate the simulations used in this work  [Table \ref{tab:SimulationList}], providing a more
detailed description of the simulations performed and their relationship to the literature.
Unless otherwise noted, we extract $\tilde{\psi}_{4,lm}(f)$ [and therefore $\tilde{h}_{lm}(f)$ and $h_{lm}(t)$] at infinity using a
perturbative extrapolation ~\cite{Nakano:2015pta} re-expressed in the Fourier domain; see \PaperPENRMethods{} for further details.

\noindent \textbf{RIT simulations}:
BBH data were evolved using the {\sc
LazEv}~\cite{Zlochower:2005bj} implementation of the moving puncture
approach~\cite{Campanelli:2005dd,Baker:2005vv} with the conformal
function $W=\sqrt{\chi}=\exp(-2\phi)$ suggested by
Ref.~\cite{Marronetti:2007wz}.  For the run presented here, we use
centered, sixth-order finite differencing in
space~\cite{Lousto:2007rj} and a fourth-order Runge Kutta time
integrator. (Note that we do not upwind the advection terms.)
This code uses the {\sc EinsteinToolkit}~\cite{Loffler:2011ay} / {\sc Cactus}~\cite{cactus_web} /
{\sc Carpet}~\cite{2004CQGra..21.1465S}
infrastructure.  The {\sc
Carpet} mesh refinement driver provides a
``moving boxes'' style of mesh refinement. In this approach, refined
grids of fixed size are arranged about the coordinate centers of both
holes.  The {\sc Carpet} code then moves these fine grids about the
computational domain by following the trajectories of the two BHs.
The RIT group used {\sc AHFinderDirect}~\cite{Thornburg2003:AH-finding} to locate
apparent horizons.  The magnitude of the horizon spin  is computed using
the {\it isolated horizon} (IH) algorithm detailed in
Ref.~\cite{Dreyer02a} and as implemented in Ref.~\cite{Campanelli:2006fy}.
Note that once we have the horizon spin, we can calculate the horizon
mass via the Christodoulou formula
\begin{equation}
{m_H} = \sqrt{m_{\rm irr}^2 + S_H^2/(4 m_{\rm irr}^2)} \,,
\end{equation}
where $m_{\rm irr} = \sqrt{A/(16 \pi)}$, $A$ is the surface area of
the horizon, and $S_H$ is the spin angular momentum of the BH (in
units of $M^2$).  

The 128 simulations reported in \citet{2015PhRvD..92b4022Z}, denoted in Table \ref{tab:SimulationList} by RIT-Kicks,
have only one black hole spinning with $|\mathbf{\chi}|=0.8$.   For a handful of these simulations, the estimate of the
final black hole mass and spin has been updated since the original publication. 

These simulations include 
(a) a  simulation with large transverse spins and several spin precession cycles which fits the data well
\cite{2015PhRvL.114n1101L}; 
(b) a wide range of simulations with large aligned and antialigned spins for mass ratios near and far from unity \cite{2014PhRvD..90j4004H};
(c) a set of  simulations with targeted mass ratios  and spins, designed to
systematically explore the parameter space and reconstruct generic recoil kicks when $q>1$
\cite{2015PhRvD..92b4022Z}; 
(d) and a set of equal mass simulations with large spins ($0.8$) and generic orientations, designed to systematically
explore the parameter space and reconstruct recoil when $q=1$ \cite{2013PhRvD..87h4027L}.

\noindent \textbf{SXS simulations}: SXS provided simulations from their public catalog -- initially reported in
\cite{2013PhRvL.111x1104M}  -- as well as several selected followup simulations.  
The SXS collaboration uses the Spectral Einstein Code (SpEC) \cite{spec} for evolution. 
Quasiequilibrium initial data are constructed in the extended conformal thin-sandwich formalism
using a pseudo-spectral elliptic solver \cite{2003CoPhC.152..253P,2008PhRvD..78h4017L}
The evolution occurs on a grid extending from inner excision boundaries,
 slightly inside the apparent horizons, to 
an outer boundary on which constraint-preserving boundary conditions
are imposed \cite{2007CQGra..24.4053R}. 
The code uses a first-order generalized harmonic
representation of Einstein's equations
with damped harmonic gauge
\cite{2002PhRvD..65d4029G,2005CQGra..22..425P,2006CQGra..23S.447L,2006PhRvD..74j4006S,2009PhRvD..80l4010S}.  
After merger, the grid is updated to include only one excision boundary
\cite{2009PhRvD..79b4003S,2013CQGra..30k5001H}.
The  excision boundaries are dynamically adjusted to conform to the shapes of the apparent horizons
\cite{2009PhRvD..79b4003S,2013CQGra..30k5001H}. 
The initial orbital eccentricity is reduced with an iterative procedure \cite{2012PhRvD..86h4033B,2011PhRvD..83j4034B}.  
Other improvements have been applied to enable long simulations \cite{2015PhRvL.115c1102S} and simulations of
highly-spinning black holes \cite{2015CQGra..32j5009S}.  

\noindent \textbf{GT simulations}: 
 Initial data was evolved with  \texttt{Maya}, which was used in previous black hole-black hole (BH-BH) studies
 \cite{Herrmann:2007ex,Herrmann:2007ac,Hinder:2007qu,Healy:2008js,Hinder:2008kv,Healy:2009zm,Healy:2009ir,Bode:2009mt}. %
The grid structure for each run consisted of  10 levels of refinement 
provided by \textsc{Carpet} \cite{2004CQGra..21.1465S}, a 
mesh refinement package for \textsc{Cactus} \cite{cactus_web}. 
Each successive level's resolution decreased by a factor of 2.
Sixth-order spatial finite differencing was used with the BSSN equations 
implemented with Kranc \cite{Husa:2004ip}. 

Simulations denoted by GT-Aligned refer to the z, zq, and zU series in
\cite{gwastro-mergers-nr-Alignment-ROS-CorotatingWaveforms,gwastro-mergers-nr-Alignment-ROS-Polarization}; the
GT-Misaligned case refers to the S and Sq series; and GT-Tilting refers to the T and  Tq series.  Where available, we
adopt the naming convention used in \cite{gr-nr-vacuum-GTcatalog-2016}. 
In particular, the  452 simulations in \cite{gr-nr-vacuum-GTcatalog-2016} survey the most extensive parameter space of
binary black hole (BBH) systems with  49
non-spinning, 81 aligned-spinning and 324 generic precessing spins BBH simulations.   They cover mass-ratios ranging
from $q\le 15$ for non-spinning and $q\le 8$ for precessing spinning BBH systems,   and include generic spin orientations and spin magnitudes, $|a| < 0.8$. 

\noindent \textbf{BAM simulations}: The Cardiff-UIB group provided \SimulationsProvidedNumberBAM{} simulations using
parameters similar to the event, with approximately random initial configurations within the 99\% credible region
inferred for \TheEvent{} in \PaperPE{}.
These BBH simulations were produced by the bifunctional adaptive mesh (BAM)
NR code \cite{PhysRevD.77.024027,husa2008}. 
The BAM code  solves the Einstein
evolutions equations using the BSSN \cite{1995PhRvD..52.5428S,1999PhRvD..59b4007B} 
formulation of the $3+1$ decomposed Einstein field equations.
The BSSN equations are integrated with a fourth order finite-difference
Runge-Kutta time integrator, with a fixed time step along with
a sixth order accurate finite difference algorithm based on the method-of-lines
for spatial derivatives.
The $\chi$ variation of the moving-puncture method is used where a new conformal
factor defined as $\chi = \psi^{-4}$ which is finite at the puncture.
The lapse and shift gauge functions are evolved using the $1+\text{Log}$ slicing
condition and the Gamma driver shift condition respectively.
Conformally flat puncture initial data \cite{Cook:1989fb,Brandt:1997tf,Bowen:1980yu}
are calculated using the pseudospectral elliptic solver described in \cite{Ansorg:2004ds}.

\subsection{Followup simulations}
Several groups performed new simulations in response to \TheEvent{}, indicated in Table \ref{tab:SimulationList} by an
asterisk (*).  Some of these simulations were made available for this analysis.  
The SXS group performed \SimulationsProvidedNumberFollowupSXS{} targeted simulations near the maximum \emph{a
  posterori} parameters reported in \PaperPE.
The RIT group  performed a systematic followup campaign on nonprecessing binaries, targeting the
mass ratio and spin range favored by \PaperPE.  This campaign included \SimulationsProvidedNumberFollowupRIT{}
new simulations of non-precessing binaries
in the range of mass ratio $1/2\leq q\leq1$ for spinning binaries and up
to $q=1/6$ without spin.  So far  the sequence also includes  11 new precessing simulations in the
observationally-relevant mass ratio range of $1/3\leq q\leq 3/4$ to further calibrate results.
\begin{longtable}{ll|lllllll|ll|clll}
\caption{\label{tab:SimulationList}\textbf{List of simulations}: Table of simulations used in this work. Columns indicate the group; an (internal) shorthand for the simulation; the mass ratio; and the components of the dimensionless spins $\mathbf{\chi}_1 = \mathbf{S}_i/m_i^2$; the effective aligned spin $\xi$; the estimated initial starting orbital frequency $M\omega_0$; and (where available) the final black mass and spin.  [We indicate where the black hole mass and spin was unavailable by using X for the corresponding entry.] (The printed table only shows a few entries from each group; the full table is available as online supplementary material.) } \\ \hline

\hline Name & Key & q & $\chi_{1,x}$ & $\chi_{1,y}$ & $\chi_{1,z}$ & $\chi_{2,x}$ & $\chi_{2,y}$ & $\chi_{2,z}$ &  $\chi_{\rm eff}$ & $M\omega_0$ & $M_{f}/M$ & $a_{f}$ \\ \hline \endfirsthead \multicolumn{4}{r}{Continued on next page}\\ \endfoot \\ \endlastfoot
{\small RIT-Generic} & {\tiny \verb|D10.50_q0.1667_a0.0_0.0_n100|}(*) & 6.000 &-&-&-&-&-& - &-& 0.025 & 0.986 & 0.372 \\
{\small RIT-Generic} & {\tiny \verb|D10_q0.33_a-0.8_xi0_n100|}(*) & 2.999 & 0.757 & 0.030 & 0.259 &-&-& -0.800 & -0.006 & 0.029 & 0.965 & 0.756 \\
{\small RIT-Generic} & {\tiny \verb|D10_q0.33_a0.8_xi0_n120|}(*) & 2.999 & 0.754 & 0.031 & -0.268 &-&-& 0.800 & -0.001 & 0.029 & 0.970 & 0.607 \\
{\small RIT-Generic} & {\tiny \verb|D10_q0.50_a-0.50_0.50_n100|}(*) & 2.000 &-&-& 0.500 &-&-& -0.500 & 0.167 & 0.028 & 0.953 & 0.751 \\
{\small RIT-Generic} & {\tiny \verb|D10_q0.50_a-0.8_xi0_n100|}(*) & 2.000 & 0.696 & 0.059 & 0.392 &-& -0.006 & -0.801 & -0.005 & 0.030 & 0.956 & 0.768 \\
{\small SXS-All} & {\tiny \verb|SXS:BBH:0001|} & 1.000 &-& - &-&-&-&-&-& 0.013 & 0.952 & 0.686 \\
{\small SXS-All} & {\tiny \verb|SXS:BBH:0010|} & 1.501 & 0.248 & 0.028 & -0.433 & - &-&-& -0.260 & 0.014 & 0.962 & 0.563 \\
{\small SXS-All} & {\tiny \verb|SXS:BBH:0100|} & 1.500 & - & - &-&-& - &-&-& 0.012 & 0.955 & 0.664 \\
{\small SXS-All} & {\tiny \verb|SXS:BBH:0101|} & 1.501 &-& - & -0.500 & - & - &-& -0.300 & 0.018 & 0.963 & 0.540 \\
{\small SXS-All} & {\tiny \verb|SXS:BBH:0102|} & 1.500 & 0.496 & 0.051 & -0.001 & 0.494 & 0.071 & -0.001 & -0.001 & 0.015 & 0.954 & 0.695 \\
{\small RIT-Kicks} & {\tiny \verb|RIT:BBH:NQ16TH115PH0|} & 6.000 & 0.725 &-& -0.338 &-&-&-& -0.290 & 0.033 & 0.991 & 0.554 \\
{\small RIT-Kicks} & {\tiny \verb|RIT:BBH:NQ16TH115PH120|} & 6.000 & -0.363 & 0.628 & -0.338 &-&-&-& -0.290 & 0.034 & 0.991 & 0.552 \\
{\small RIT-Kicks} & {\tiny \verb|RIT:BBH:NQ16TH115PH150|} & 6.000 & -0.628 & 0.363 & -0.338 &-&-&-& -0.290 & 0.034 & 0.991 & 0.556 \\
{\small RIT-Kicks} & {\tiny \verb|RIT:BBH:NQ16TH115PH30|} & 6.000 & 0.628 & 0.363 & -0.338 &-&-&-& -0.290 & 0.032 & 0.991 & 0.553 \\
{\small RIT-Kicks} & {\tiny \verb|RIT:BBH:NQ16TH115PH60|} & 6.000 & 0.363 & 0.628 & -0.338 &-&-&-& -0.290 & 0.034 & 0.991 & 0.556 \\
{\small RIT-OlderWork} & {\tiny \verb|RIT:BBH:KTH22.5PH0|} & 1.000 & -0.026 & 0.304 & 0.760 & -0.008 & 0.310 & -0.759 & 0.001 & 0.042 & 0.960 & 0.695 \\
{\small RIT-OlderWork} & {\tiny \verb|RIT:BBH:KTH22.5PH120|} & 1.000 & -0.272 & -0.157 & 0.757 & -0.272 & -0.157 & -0.757 & - & 0.043 & 0.961 & 0.698 \\
{\small RIT-OlderWork} & {\tiny \verb|RIT:BBH:KTH22.5PH150|} & 1.000 & -0.157 & -0.272 & 0.757 & -0.157 & -0.272 & -0.757 & - & 0.043 & 0.961 & 0.697 \\
{\small RIT-OlderWork} & {\tiny \verb|RIT:BBH:KTH22.5PH30|} & 1.000 & -0.185 & 0.257 & 0.756 & -0.157 & 0.272 & -0.757 & -0.001 & 0.042 & 0.960 & 0.695 \\
{\small RIT-OlderWork} & {\tiny \verb|RIT:BBH:KTH22.5PH60|} & 1.000 & -0.297 & 0.138 & 0.751 & -0.272 & 0.157 & -0.757 & -0.003 & 0.042 & 0.960 & 0.695 \\
{\small GT} & {\tiny \verb|GT:BBH:564|} & 1.000 &-&-& -0.400 &-&-& -0.400 & -0.400 & 0.026 & 0.961 & 0.560 \\
{\small GT} & {\tiny \verb|GT:BBH:476|} & 1.000 &-&-& -0.200 &-&-& -0.200 & -0.200 & 0.025 & 0.956 & 0.624 \\
{\small GT} & {\tiny \verb|(0.0,1.0)|} & 1.000 &-&-&-&-&-&-&-& 0.030 & 0.952 & 0.686 \\
{\small GT} & {\tiny \verb|(0,1.0,'M100')|} & 1.000 &-&-&-&-&-&-&-& 0.029 & 0.951 & 0.687 \\
{\small GT} & {\tiny \verb|(0.0,1.0,'M120','D11')|} & 1.000 &-&-&-&-&-&-&-& 0.029 & 0.951 & 0.686 \\
{\small GT} & {\tiny \verb|GT:BBH:456|} & 1.500 & 0.346 &-& 0.200 &-&-& 0.400 & 0.280 & 0.024 & 0.947 & 0.753 \\
{\small GT} & {\tiny \verb|GT:BBH:455|} & 1.500 & 0.424 &-& 0.424 &-&-& 0.600 & 0.495 & 0.020 & 0.937 & 0.822 \\
{\small GT} & {\tiny \verb|GT:BBH:457|} & 1.500 & 0.520 &-& 0.300 &-&-& 0.600 & 0.420 & 0.021 & X & X \\
{\small GT} & {\tiny \verb|GT:BBH:764|} & 1.500 & 0.600 &-&-&-&-& 0.600 & 0.240 & 0.021 & X & X \\
{\small GT} & {\tiny \verb|GT:BBH:458|} & 2.000 & 0.346 &-& 0.200 &-&-& 0.400 & 0.267 & 0.023 & 0.954 & 0.722 \\
{\small GT} & {\tiny \verb|GT:BBH:550|} & 2.000 & 0.424 &-& -0.424 &-&-& 0.600 & -0.083 & 0.032 & 0.964 & 0.549 \\
{\small GT} & {\tiny \verb|GT:BBH:545|} & 2.000 &-&-& -0.600 &-&-& 0.600 & -0.200 & 0.033 & 0.967 & 0.465 \\
{\small GT} & {\tiny \verb|GT:BBH:556|} & 2.000 & -0.600 &-& - &-&-& 0.600 & 0.200 & 0.032 & 0.955 & 0.698 \\
{\small GT} & {\tiny \verb|EK_D6.2_a0.6_th000_M77|} & 1.000 & 0.584 & 0.143 & 0.002 & -0.584 & -0.143 & 0.002 & 0.002 & 0.070 & 0.951 & 0.686 \\
{\small GT} & {\tiny \verb|GT:BBH:482|} & 1.000 & 0.520 & 0.300 &-& -0.520 & -0.300 &-&-& 0.071 & 0.951 & 0.684 \\
{\small GT} & {\tiny \verb|GT:BBH:483|} & 1.000 & 0.424 & 0.424 &-& -0.424 & -0.424 &-&-& 0.071 & 0.950 & 0.683 \\
{\small GT} & {\tiny \verb|GT:BBH:484|} & 1.000 & 0.300 & 0.520 &-& -0.300 & -0.520 &-&-& 0.072 & 0.950 & 0.681 \\
{\small GT} & {\tiny \verb|GT:BBH:485|} & 1.000 &-& 0.600 &-& - & -0.600 &-&-& 0.072 & 0.949 & 0.680 \\
{\small GT} & {\tiny \verb|aa_b5_a0.2_M77|} & 1.000 &-&-& 0.200 &-&-& -0.200 &-& 0.027 & X & X \\
{\small GT} & {\tiny \verb|aa_b5_a0.4_M77|} & 1.000 &-&-& 0.400 &-&-& -0.400 &-& 0.028 & X & X \\
{\small GT} & {\tiny \verb|aa_b5_a0.6_M77|} & 1.000 &-&-& 0.600 &-&-& -0.600 &-& 0.029 & X & X \\
{\small GT} & {\tiny \verb|aa_b5_a0.8_M77|} & 1.000 &-&-& 0.800 &-&-& -0.800 &-& 0.031 & X & X \\
{\small GT} & {\tiny \verb|fr_b5_a0.6_random2_M77|} & 1.000 &-&-& 0.600 &-&-& -0.600 &-& 0.027 & X & X \\
{\small GT} & {\tiny \verb|fr_b3.1_a0.6_oth.000_M77|} & 1.000 &-&-& 0.600 & -0.600 &-&-& 0.300 & 0.057 & X & X \\
{\small GT} & {\tiny \verb|fr_b3.1_a0.6_oth.015_M77|} & 1.000 & 0.155 &-& 0.580 & -0.600 &-&-& 0.290 & 0.057 & X & X \\
{\small GT} & {\tiny \verb|fr_b3.1_a0.6_oth.030_M77|} & 1.000 & 0.300 &-& 0.520 & -0.600 &-&-& 0.260 & 0.058 & X & X \\
{\small GT} & {\tiny \verb|fr_b3.1_a0.6_oth.045_M77|} & 1.000 & 0.424 &-& 0.424 & -0.600 &-&-& 0.212 & 0.058 & X & X \\
{\small GT} & {\tiny \verb|fr_b3.1_a0.6_oth.060_M77|} & 1.000 & 0.520 &-& 0.300 & -0.600 &-&-& 0.150 & 0.059 & X & X \\
{\small GT} & {\tiny \verb|D10_q7.00_a0.0_m320|} & 7.000 &-&-&-&-&-&-&-&-& X & X \\
{\small GT} & {\tiny \verb|GT:BBH:860|} & 1.000 & 0.109 & 0.481 & 0.342 & 0.460 & -0.287 & 0.257 & 0.299 & 0.042 & 0.940 & 0.783 \\
{\small GT} & {\tiny \verb|GT:BBH:861|} & 1.000 & -0.159 & -0.414 & -0.404 & 0.297 & 0.521 & -0.020 & -0.212 & 0.049 & 0.957 & 0.620 \\
{\small GT} & {\tiny \verb|GT:BBH:862|} & 1.000 & 0.542 & -0.255 & -0.034 & 0.053 & 0.188 & -0.567 & -0.301 & 0.053 & 0.958 & 0.607 \\
{\small GT} & {\tiny \verb|GT:BBH:863|} & 1.000 & -0.512 & 0.270 & -0.157 & -0.506 & -0.175 & -0.270 & -0.214 & 0.053 & 0.956 & 0.663 \\
{\small BAM-GitAnnex} & {\tiny \verb|BAM150914:31|}(*) & 1.200 & 0.384 & -0.135 & -0.119 & -0.354 & 0.218 & 0.086 & -0.026 & 0.026 & X & X \\
{\small BAM-GitAnnex} & {\tiny \verb|BAM150914:1|}(*) & 1.200 & 0.384 & -0.135 & -0.119 & -0.354 & 0.218 & 0.086 & -0.026 & 0.025 & X & X \\
{\small BAM-GitAnnex} & {\tiny \verb|BAM150914:29|}(*) & 1.200 & 0.123 & 0.366 & -0.175 & 0.136 & -0.460 & 0.469 & 0.118 & 0.027 & X & X \\
{\small BAM-GitAnnex} & {\tiny \verb|BAM150914:18|}(*) & 1.200 & -0.161 & -0.207 & 0.145 & 0.378 & 0.352 & 0.428 & 0.274 & 0.023 & X & X \\
{\small BAM-GitAnnex} & {\tiny \verb|BAM150914:26|}(*) & 1.200 & -0.095 & 0.404 & -0.088 & 0.605 & -0.442 & 0.399 & 0.133 & 0.029 & X & X \\
\end{longtable}

\section{Tables I: Rankings}
In this section, we enumerate the simulations used in this work, ordered by one measure of
their similarity with the data [ $\ln L$, in Table \ref{tab:SimulationRanks:30}].  For nonprecessing binaries, Figure
\ref{fig:ap:AlignedSpinMaxLmarg} provides a visual illustration of some trends in $\ln L$ versus mass ratio and the two
component spins.

\begin{longtable}{l|l|lrrrrrr|rr|r}
\caption{\label{tab:SimulationRanks:30}\textbf{Peak Marginalized $\ln L$ I: Consistency between simulations}: Peak value of the marginalized log likelihood $\ln L$ [Eq. (\ref{eq:lnL})] evaluated using a lower frequency $f_{\rm low}=30\unit{Hz}$ and all modes with $l \le 2$; the simulation key, described in Table \ref{tab:SimulationList} [an asterisk (*) denotes a new simulation motivated by \TheEvent{}, and a (+) denotes one of the simulations reported in \PaperDetection];  the \emph{initial} spins of the simulation (using $-$ to denote zero, to enhance readability); the initial $\chi_{\rm eff}$; the total (redshifted) mass of the best fit; and the starting frequency (in Hz) of the best fit.  Though omitting information accessible to the longest simulations, this choice of low-frequency cutoff eliminates systematic biases associated with simulation duration, which differs across our archive, as seen by the last column.  } \\ \hline \hline $\ln L$ & Key & $q$ & $\chi_{1,x}$ & $\chi_{1,y}$ & $\chi_{1,z}$ & $\chi_{2,x}$ & $\chi_{2,y}$ & $\chi_{2,z}$ & $\chi_{\rm eff}$ & $M_z/M_\odot$  & $f_{\rm start} ({\rm Hz})$ \\ \hline \endfirsthead \multicolumn{4}{r}{Continued on next page}\\ \endfoot \\ \endlastfoot 
272.2 & \verb|SXS:BBH:0310|(*) & 1.221 & - &-&-& - & - &-& 0.00 & 73.0 & 15.3 \\
272.1 & \verb|D12_q1.00_a-0.25_0.25_n100|(*) & 1.0 &-&-& 0.250 &-&-& -0.250 & -0.00 & 73.2 & 20.5 \\
272.1 & \verb|SXS:BBH:0002| & 1.0 &-& - &-&-&-&-& 0.00 & 73.2 & 9.9 \\
271.8 & \verb|D11_q0.75_a0.0_0.0_n100|(*) & 1.333 &-&-& - &-&-& - & -0.00 & 72.1 & 23.1 \\
271.8 & \verb|SXS:BBH:0305|(*+) & 1.221 & - &-& 0.330 & - & - & -0.440 & -0.02 & 74.2 & 15.3 \\
271.6 & \verb|SXS:BBH:0218| & 1.0 &-& - & -0.500 & - & - & 0.500 & 0.00 & 73.3 & 10.7 \\
271.6 & \verb|SXS:BBH:0198| & 1.202 & - & - & - & - & - &-& 0.00 & 73.4 & 12.9 \\
271.6 & \verb|SXS:BBH:0307|(*) & 1.228 & - & - & 0.320 &-& - & -0.580 & -0.08 & 70.0 & 17.6 \\
271.6 & \verb|GT:BBH:476| & 1.0 &-&-& -0.200 &-&-& -0.200 & -0.20 & 67.9 & 24.3 \\
271.6 & \verb|S0_D10.04_q1.3333_a0.45_-0.80_n100| & 1.334 &-&-& 0.450 &-&-& -0.801 & -0.09 & 71.9 & 27.9 \\
271.5 & \verb|D12.00_q0.85_a0.0_0.0_n100|(*) & 1.176 &-&-& - &-&-& - & -0.00 & 73.0 & 20.6 \\
271.5 & \verb|D12.25_q0.82_a-0.44_0.33_n100|(*+) & 1.22 &-&-& 0.330 &-&-& -0.440 & -0.02 & 72.9 & 20.2 \\
271.5 & \verb|SXS:BBH:0312|(*) & 1.203 &-& - & 0.390 &-&-& -0.480 & -0.00 & 73.9 & 15.2 \\
271.4 & \verb|SXS:BBH:0127| & 1.34 & 0.010 & -0.077 & -0.017 & -0.061 & -0.065 & -0.179 & -0.09 & 71.5 & 14.7 \\
271.4 & \verb|SXS:BBH:0115| & 1.07 & 0.019 & 0.013 & -0.204 & 0.243 & -0.067 & 0.291 & 0.04 & 74.1 & 14.6 \\
271.3 & \verb|SXS:BBH:0213| & 1.0 &-& - & -0.800 & - &-& 0.800 & 0.00 & 73.2 & 12.4 \\
271.3 & \verb|UD_D10.01_q1.00_a0.4_n100| & 1.0 &-&-& 0.400 &-&-& -0.400 & -0.00 & 73.4 & 26.7 \\
271.2 & \verb|D12_q1.00_a-0.25_0.00_n100|(*) & 1.0 &-&-& - &-&-& -0.250 & -0.12 & 69.4 & 21.8 \\
271.2 & \verb|SXS:BBH:0222| & 1.0 &-&-& -0.300 &-&-&-& -0.15 & 69.1 & 12.5 \\
271.2 & \verb|SXS:BBH:0217| & 1.0 &-& - & -0.600 & - &-& 0.600 & 0.00 & 73.2 & 12.4 \\
271.1 & \verb|D10_q0.75_a-0.5_0.25_n100|(*) & 1.333 &-&-& 0.250 &-&-& -0.500 & -0.07 & 71.9 & 27.4 \\
271.0 & \verb|BAM150914:24|(*) & 1.2 & 0.151 & 0.396 & 0.017 & -0.278 & -0.605 & -0.085 & -0.03 & 72.2 & 17.8 \\
270.9 & \verb|GW15_D12_q1.22_a0.33_-0.44_m140|(*) & 1.22 & - &-& 0.330 & - &-& -0.440 & -0.02 & 71.7 & 19.7 \\
270.9 & \verb|SXS:BBH:0308[Lev3]|(*) & 1.228 & 0.072 & 0.072 & 0.325 & 0.201 & 0.285 & -0.571 & -0.08 & 70.5 & 17.7 \\
270.9 & \verb|SXS:BBH:0120| & 1.12 & 0.138 & -0.200 & -0.008 & -0.065 & -0.241 & -0.099 & -0.05 & 70.3 & 14.8 \\
270.8 & \verb|SXS:BBH:0006| & 1.345 & 0.234 & 0.148 & -0.161 & 0.091 & 0.064 & -0.101 & -0.14 & 69.6 & 13.6 \\
270.8 & \verb|SXS:BBH:0313|(*) & 1.217 &-& - & 0.380 &-&-& -0.520 & -0.03 & 72.5 & 15.6 \\
270.7 & \verb|GT:BBH:370| & 1.15 &-&-&-&-&-&-& 0.00 & 74.0 & 25.8 \\
270.7 & \verb|SXS:BBH:0308| & 1.228 & 0.094 & 0.056 & 0.322 & 0.266 & 0.213 & -0.576 & -0.08 & 71.8 & 17.2 \\
270.6 & \verb|SXS:BBH:0123| & 1.1 & 0.267 & 0.020 & -0.415 & 0.038 & -0.054 & 0.126 & -0.16 & 68.0 & 16.3 \\
270.6 & \verb|D11_q0.75_a-0.5_0.5_n100|(*) & 1.333 &-&-& 0.500 &-&-& -0.500 & 0.07 & 76.7 & 21.6 \\
270.5 & \verb|SXS:BBH:0129| & 1.36 & -0.001 & - & 0.088 & 0.193 & -0.289 & -0.075 & 0.02 & 74.0 & 14.0 \\
270.5 & \verb|SXS:BBH:0117| & 1.08 & 0.118 & -0.069 & 0.070 & -0.302 & -0.298 & -0.200 & -0.06 & 72.8 & 14.7 \\
270.4 & \verb|SXS:BBH:0003| & 1.0 & 0.497 & 0.053 & - &-&-&-& -0.00 & 72.5 & 11.6 \\
270.4 & \verb|D10_q0.75_a-0.25_0.25_n100|(*) & 1.333 &-&-& 0.250 &-&-& -0.250 & 0.04 & 75.7 & 24.7 \\
270.4 & \verb|SXS:BBH:0224| & 1.0 & - &-& 0.400 & - & - & -0.800 & -0.20 & 67.3 & 13.0 \\
270.3 & \verb|D11_q0.75_a-0.8500_0.6375_n100| & 1.334 &-&-& 0.638 & 0.001 & 0.003 & -0.851 & -0.00 & 74.8 & 22.6 \\
270.3 & \verb|U0_D9.53_q1.00_a0.0_n100| & 1.0 &-&-& - &-&-& - & -0.00 & 73.7 & 28.6 \\
270.2 & \verb|SXS:BBH:0211| & 1.0 &-&-& -0.900 &-&-& 0.900 & 0.00 & 73.5 & 12.3 \\
270.2 & \verb|SXS:BBH:0116| & 1.08 & -0.078 & 0.065 & 0.033 & 0.185 & 0.007 & 0.103 & 0.07 & 76.4 & 13.7 \\
270.2 & \verb|D11_q0.75_a0.5_-0.5_n100|(*) & 1.333 &-&-& -0.500 &-&-& 0.500 & -0.07 & 70.2 & 23.9 \\
270.2 & \verb|GW15_D12_q1.19_a0.42_-0.38_m140|(*) & 1.19 & - &-& 0.420 & - &-& -0.380 & 0.05 & 74.0 & 18.6 \\
270.1 & \verb|GT:BBH:900|(*) & 1.2 & 0.400 &-&-& 0.400 &-&-& 0.00 & 74.3 & 21.8 \\
270.1 & \verb|D11_q0.75_a-0.5_0.0_n100|(*) & 1.333 &-&-& - &-&-& -0.500 & -0.21 & 67.8 & 24.4 \\
270.0 & \verb|GT:BBH:898|(*) & 1.2 &-&-&-&-&-&-& 0.00 & 74.2 & 18.0 \\
269.9 & \verb|aa_b5_a0.6_M77| & 1 &-&-& 0.600 &-&-& -0.600 & 0.00 & 73.7 & 25.2 \\
269.9 & \verb|SXS:BBH:0125| & 1.27 & 0.012 & 0.045 & -0.058 & 0.389 & 0.241 & 0.070 & -0.00 & 75.1 & 13.3 \\
269.9 & \verb|D21.5_q1_a0.2_0.8_th104.4775_n100| & 1.001 &-&-& 0.200 & 0.775 &-& -0.200 & -0.00 & 74.2 & 9.7 \\
269.9 & \verb|SXS:BBH:0131| & 1.55 & 0.042 & -0.014 & -0.070 & 0.105 & 0.017 & -0.175 & -0.11 & 71.0 & 14.8 \\
269.8 & \verb|SXS:BBH:0096| & 1.501 & 0.497 & 0.051 & - &-&-&-& -0.00 & 73.0 & 13.2 \\
269.8 & \verb|SXS:BBH:0088| & 1.0 & 0.495 & 0.067 &-&-&-&-& 0.00 & 74.3 & 11.6 \\
269.7 & \verb|D10_q0.75_a-0.8_xi0_n100|(*) & 1.333 & 0.538 & 0.056 & 0.590 &-&-& -0.801 & -0.01 & 75.2 & 26.5 \\
269.7 & \verb|SXS:BBH:0029| & 1.5 & 0.496 & 0.051 & -0.001 & 0.494 & 0.070 & -0.001 & -0.00 & 74.3 & 13.4 \\
269.6 & \verb|BAM150914:25|(*) & 1.2 & 0.119 & -0.407 & 0.017 & 0.125 & 0.656 & -0.064 & -0.02 & 72.3 & 22.3 \\
269.6 & \verb|SXS:BBH:0163| & 1.0 & 0.441 & 0.290 & -0.284 & 0.424 & 0.266 & 0.331 & 0.02 & 73.5 & 13.6 \\
269.6 & \verb|SXS:BBH:0226| & 1.0 &-&-& 0.500 & - &-& -0.900 & -0.20 & 67.4 & 12.9 \\
269.5 & \verb|D10_q0.75_a0.25_-0.25_n100|(*) & 1.333 &-&-& -0.250 &-&-& 0.250 & -0.04 & 71.8 & 26.2 \\
269.5 & \verb|BAM150914:9|(*) & 1.2 & 0.554 & -0.314 & 0.212 & 0.008 & 0.643 & -0.191 & 0.03 & 72.7 & 27.9 \\
269.4 & \verb|SXS:BBH:0316|(*) & 1.186 & 0.241 & 0.170 & 0.299 & -0.203 & -0.172 & -0.271 & 0.04 & 75.2 & 15.0 \\
269.4 & \verb|SXS:BBH:0121| & 1.12 & -0.061 & -0.109 & 0.356 & -0.323 & -0.127 & -0.297 & 0.05 & 76.5 & 14.7 \\
269.4 & \verb|SXS:BBH:0097| & 1.501 & 0.495 & 0.065 & 0.001 &-&-&-& 0.00 & 72.7 & 15.0 \\
269.4 & \verb|BAM150914:31|(*) & 1.2 & 0.384 & -0.135 & -0.119 & -0.354 & 0.218 & 0.086 & -0.03 & 71.5 & 23.9 \\
269.4 & \verb|SXS:BBH:0100| & 1.5 & - & - &-&-& - &-& 0.00 & 74.4 & 10.7 \\
269.2 & \verb|D10_q0.75_a0.5_-0.25_n100|(*) & 1.333 &-&-& -0.250 &-&-& 0.500 & 0.07 & 75.2 & 24.7 \\
269.2 & \verb|GT:BBH:448| & 1.0 &-&-&-&-&-&-& 0.00 & 76.0 & 21.5 \\
269.1 & \verb|SXS:BBH:0135| & 1.64 & -0.110 & 0.027 & 0.024 & 0.211 & -0.144 & -0.229 & -0.07 & 71.5 & 14.5 \\
269.1 & \verb|SXS:BBH:0149| & 1.0 &-& - & -0.200 & - &-& -0.200 & -0.20 & 68.3 & 14.8 \\
269.0 & \verb|BAM150914:6|(*) & 1.2 & 0.662 & -0.070 & 0.083 & -0.358 & 0.201 & -0.106 & -0.00 & 72.7 & 23.8 \\
269.0 & \verb|BAM150914:14|(*) & 1.2 & 0.662 & -0.015 & 0.106 & -0.384 & 0.113 & -0.141 & -0.01 & 71.0 & 23.9 \\
268.9 & \verb|RIT:BBH:STH45PH30| & 1.0 & -0.337 & 0.463 & 0.585 & 0.290 & -0.502 & -0.579 & 0.00 & 71.0 & 37.6 \\
268.9 & \verb|SXS:BBH:0147| & 1.0 & 0.404 & 0.294 & -0.001 & -0.404 & -0.294 & -0.001 & -0.00 & 71.0 & 24.6 \\
268.9 & \verb|SXS:BBH:0194| & 1.518 &-& - &-& - &-& - & -0.00 & 73.0 & 13.9 \\
268.9 & \verb|GT:BBH:717| & 1.1 &-&-&-&-&-&-& 0.00 & 74.0 & 31.3 \\
268.8 & \verb|D11_q0.75_a0.8_0.4_PNr500_th1d_n100|(*) & 1.334 & 0.074 & -0.374 & 0.123 & -0.420 & -0.531 & -0.427 & -0.11 & 71.8 & 23.5 \\
268.8 & \verb|SXS:BBH:0138| & 1.7 & -0.044 & 0.425 & 0.042 & -0.012 & -0.008 & -0.111 & -0.01 & 73.9 & 14.1 \\
268.8 & \verb|SXS:BBH:0119| & 1.12 & -0.012 & 0.068 & 0.260 & 0.078 & -0.003 & 0.006 & 0.14 & 78.8 & 13.4 \\
268.6 & \verb|BAM150914:15|(*) & 1.2 & 0.276 & -0.106 & 0.052 & 0.144 & 0.268 & 0.295 & 0.16 & 77.4 & 18.4 \\
268.6 & \verb|SXS:BBH:0133| & 1.63 & 0.098 & 0.042 & -0.134 & -0.107 & -0.110 & -0.021 & -0.09 & 71.7 & 14.0 \\
268.6 & \verb|BAM150914:2|(*) & 1.2 & -0.099 & -0.377 & -0.167 & -0.108 & 0.484 & 0.452 & 0.11 & 76.3 & 21.6 \\
268.5 & \verb|SXS:BBH:0098| & 1.501 & 0.486 & 0.114 & 0.002 &-&-&-& 0.00 & 75.1 & 11.5 \\
268.4 & \verb|aa_b5_a0.8_M77| & 1 &-&-& 0.800 &-&-& -0.800 & 0.00 & 75.3 & 26.5 \\
268.4 & \verb|GT:BBH:717| & 1.1 &-&-&-&-&-&-& 0.00 & 73.0 & 30.8 \\
268.4 & \verb|D11.50_q0.60_a0.0_0.0_n100|(*) & 1.667 &-&-& - &-&-& - & -0.00 & 73.6 & 21.2 \\
268.4 & \verb|d0_D10.52_q1.3333_a-0.25_n100| & 1.333 &-&-& -0.250 &-&-& - & -0.14 & 69.1 & 26.2 \\
268.3 & \verb|SXS:BBH:0223| & 1.0 &-&-& 0.300 & - &-&-& 0.15 & 78.8 & 11.7 \\
268.3 & \verb|SXS:BBH:0082| & 1.501 & 0.496 & 0.053 & - &-&-&-& -0.00 & 75.5 & 13.7 \\
268.3 & \verb|RIT:BBH:STH45PH60| & 1.0 & -0.537 & 0.242 & 0.570 & 0.502 & -0.290 & -0.579 & -0.00 & 68.9 & 37.9 \\
268.3 & \verb|SXS:BBH:0027| & 1.5 & 0.497 & 0.051 & - & -0.494 & -0.071 &-& 0.00 & 72.8 & 13.9 \\
268.2 & \verb|BAM150914:5|(*) & 1.2 & 0.649 & 0.149 & 0.086 & -0.385 & -0.130 & -0.124 & -0.01 & 71.0 & 22.1 \\
268.2 & \verb|D12_q1.00_a-0.25_-0.25_n100|(*) & 1.0 &-&-& -0.250 &-&-& -0.250 & -0.25 & 65.7 & 23.1 \\
268.2 & \verb|aa_b5_a0.2_M77| & 1 &-&-& 0.200 &-&-& -0.200 & 0.00 & 75.3 & 23.2 \\
268.2 & \verb|SXS:BBH:0173| & 1.5 & 0.235 & 0.146 & -0.161 & 0.091 & 0.065 & -0.101 & -0.14 & 68.9 & 13.8 \\
268.2 & \verb|aa_b5_a0.4_M77| & 1 &-&-& 0.400 &-&-& -0.400 & 0.00 & 75.3 & 23.8 \\
268.1 & \verb|RIT:BBH:NTH120PH150| & 1.005 & -0.322 & -0.617 & -0.422 & - & - &-& -0.21 & 65.2 & 36.6 \\
268.1 & \verb|SXS:BBH:0103| & 1.501 & 0.496 & 0.058 &-&-&-&-& 0.00 & 75.5 & 10.5 \\
268.1 & \verb|GT:BBH:885| & 1.0 & 0.424 &-& 0.424 & -0.424 &-& -0.424 & 0.00 & 72.8 & 31.2 \\
268.1 & \verb|SXS:BBH:0004| & 1.0 & - & - & -0.500 & - & - &-& -0.25 & 66.2 & 11.3 \\
268.1 & \verb|SXS:BBH:0023| & 1.501 & 0.497 & 0.051 & 0.001 & 0.077 & -0.489 &-& 0.00 & 73.0 & 13.5 \\
268.0 & \verb|BAM150914:3|(*) & 1.2 & 0.299 & 0.028 & 0.008 & 0.192 & 0.092 & 0.367 & 0.17 & 77.3 & 18.2 \\
268.0 & \verb|BAM150914:4|(*) & 1.2 & 0.159 & -0.393 & -0.016 & 0.149 & 0.654 & -0.014 & -0.02 & 70.1 & 23.0 \\
268.0 & \verb|SXS:BBH:0021| & 1.5 & 0.496 & 0.053 & - & - & 0.001 & -0.499 & -0.20 & 70.3 & 12.7 \\
268.0 & \verb|SXS:BBH:0015| & 1.501 & 0.487 & 0.110 & 0.001 &-&-&-& 0.00 & 75.0 & 10.4 \\
\end{longtable}

\begin{longtable}{l|l|lrrrrrr|rr|r}
\caption{\label{tab:SimulationRanks:30:Lmax3}\textbf{Peak Marginalized $\ln L$ I: Consistency between simulations}: Peak value of the marginalized log likelihood $\ln L$ evaluated using a lower frequency $f_{low}=30\unit{Hz}$ [Eq. (\ref{eq:lnL})]  and all modes with $l \le 3$; the simulation key, described in Table \ref{tab:SimulationList} [an asterisk (*) denotes a new simulation motivated by \TheEvent{}]; the \emph{initial} spins of the simulation (using $-$ to denote zero, to enhance readability); the initial $\chi_{\rm eff}$; the total (redshifted) mass of the best fit; and the starting frequency (in Hz) of the best fit.   Though omitting information accessible to the longest simulations, this choice of low-frequency cutoff eliminates systematic biases associated with simulation duration, which differs across our archive.  } \\ \hline \hline $\ln L$ & Key & $q$ & $\chi_{1,x}$ & $\chi_{1,y}$ & $\chi_{1,z}$ & $\chi_{2,x}$ & $\chi_{2,y}$ & $\chi_{2,z}$ & $\chi_{\rm eff}$ & $M_z/M_\odot$  & $f_{\rm start} (Hz)$  \\ \hline \endfirsthead \multicolumn{4}{r}{Continued on next page}\\ \endfoot \\ \endlastfoot 
272.0 & \verb|SXS:BBH:0307|(*) & 1.228 & - & - & 0.320 &-& - & -0.580 & -0.08 & 70.9 & 17.3 \\
271.8 & \verb|SXS:BBH:0115| & 1.07 & 0.019 & 0.013 & -0.204 & 0.243 & -0.067 & 0.291 & 0.04 & 74.0 & 14.6 \\
271.8 & \verb|SXS:BBH:0305|(*+) & 1.221 & - &-& 0.330 & - & - & -0.440 & -0.02 & 73.5 & 15.4 \\
271.8 & \verb|D12_q1.00_a-0.25_0.25_n100|(*) & 1.0 &-&-& 0.250 &-&-& -0.250 & -0.00 & 73.6 & 20.4 \\
271.5 & \verb|SXS:BBH:0310|(*) & 1.221 & - &-&-& - & - &-& 0.00 & 72.4 & 15.4 \\
271.0 & \verb|GW15_D12_q1.22_a0.33_-0.44_m140|(*) & 1.22 & - &-& 0.330 & - &-& -0.440 & -0.02 & 72.8 & 19.4 \\
270.8 & \verb|SXS:BBH:0116| & 1.08 & -0.078 & 0.065 & 0.033 & 0.185 & 0.007 & 0.103 & 0.07 & 75.8 & 13.9 \\
270.7 & \verb|SXS:BBH:0002| & 1.0 &-& - &-&-&-&-& 0.00 & 75.1 & 9.7 \\
270.6 & \verb|SXS:BBH:0316|(*) & 1.186 & 0.241 & 0.170 & 0.299 & -0.203 & -0.172 & -0.271 & 0.04 & 74.1 & 15.2 \\
270.4 & \verb|D12.00_q0.85_a0.0_0.0_n100|(*) & 1.176 &-&-& - &-&-& - & -0.00 & 74.5 & 20.2 \\
270.4 & \verb|SXS:BBH:0120| & 1.12 & 0.138 & -0.200 & -0.008 & -0.065 & -0.241 & -0.099 & -0.05 & 70.0 & 14.9 \\
270.4 & \verb|SXS:BBH:0198| & 1.202 & - & - & - & - & - &-& 0.00 & 74.7 & 12.7 \\
270.3 & \verb|SXS:BBH:0125| & 1.27 & 0.012 & 0.045 & -0.058 & 0.389 & 0.241 & 0.070 & -0.00 & 73.5 & 13.6 \\
270.3 & \verb|GT:BBH:900|(*) & 1.2 & 0.400 &-&-& 0.400 &-&-& 0.00 & 72.7 & 22.3 \\
270.2 & \verb|D10_q0.75_a-0.5_0.25_n100|(*) & 1.333 &-&-& 0.250 &-&-& -0.500 & -0.07 & 73.7 & 26.7 \\
270.1 & \verb|GT:BBH:898|(*) & 1.2 &-&-&-&-&-&-& 0.00 & 74.8 & 17.9 \\
\end{longtable}

\begin{longtable}{l|l|lrrrrrr|rr|r}
\caption{\label{tab:SimulationRanks:AllSignal}\textbf{Peak Marginalized $\ln L$: Low frequency included}: Peak value of the marginalized log likelihood $\ln L$ evaluated using a lower frequency $f_{low}=10\unit{Hz}$ and all modes with $l \le 2$; the simulation key, described in Table \ref{tab:SimulationList} [an asterisk (*) denotes a new simulation motivated by \TheEvent{}, and a (+) denotes one of the simulations reported in \PaperDetection];  the \emph{initial} spins of the simulation (using $-$ to denote zero, to enhance readability); the initial $\xi$; and the total (redshifted) mass of the best fit.   This choice of low-frequency cutoff ensures that long simulations can make the best use of low-frequency information in the data, significantly improving our constraints on $M$ and spin precession.  } \\ \hline \hline $\ln L$ & Key & $q$ & $\chi_{1,x}$ & $\chi_{1,y}$ & $\chi_{1,z}$ & $\chi_{2,x}$ & $\chi_{2,y}$ & $\chi_{2,z}$ & $\chi_{\rm eff}$ & $M_z/M_\odot$  \\ \hline \endfirsthead \multicolumn{4}{r}{Continued on next page}\\ \endfoot \\ \endlastfoot 
277.3 & \verb|SXS:BBH:0002| & 1.0 &-& - &-&-&-&-& 0.00 & 72.7 & 10.0 \\
277.2 & \verb|SXS:BBH:0313|(*) & 1.217 &-& - & 0.380 &-&-& -0.520 & -0.03 & 72.9 & 15.5 \\
276.9 & \verb|SXS:BBH:0305|(*+) & 1.221 & - &-& 0.330 & - & - & -0.440 & -0.02 & 73.2 & 15.5 \\
276.6 & \verb|D11_q0.75_a0.0_0.0_n100|(*) & 1.333 &-&-& - &-&-& - & -0.00 & 72.6 & 22.9 \\
276.1 & \verb|SXS:BBH:0006| & 1.345 & 0.234 & 0.148 & -0.161 & 0.091 & 0.064 & -0.101 & -0.14 & 69.0 & 13.7 \\
275.4 & \verb|SXS:BBH:0096| & 1.501 & 0.497 & 0.051 & - &-&-&-& -0.00 & 72.7 & 13.3 \\
275.4 & \verb|SXS:BBH:0163| & 1.0 & 0.441 & 0.290 & -0.284 & 0.424 & 0.266 & 0.331 & 0.02 & 74.4 & 13.4 \\
275.2 & \verb|SXS:BBH:0131| & 1.55 & 0.042 & -0.014 & -0.070 & 0.105 & 0.017 & -0.175 & -0.11 & 70.8 & 14.8 \\
275.0 & \verb|GT:BBH:898|(*) & 1.2 &-&-&-&-&-&-& 0.00 & 74.2 & 18.0 \\
274.9 & \verb|SXS:BBH:0029| & 1.5 & 0.496 & 0.051 & -0.001 & 0.494 & 0.070 & -0.001 & -0.00 & 74.2 & 13.4 \\
274.8 & \verb|SXS:BBH:0100| & 1.5 & - & - &-&-& - &-& 0.00 & 72.9 & 11.0 \\
274.5 & \verb|SXS:BBH:0121| & 1.12 & -0.061 & -0.109 & 0.356 & -0.323 & -0.127 & -0.297 & 0.05 & 74.6 & 15.0 \\
274.1 & \verb|SXS:BBH:0117| & 1.08 & 0.118 & -0.069 & 0.070 & -0.302 & -0.298 & -0.200 & -0.06 & 72.5 & 14.8 \\
274.1 & \verb|SXS:BBH:0316|(*) & 1.186 & 0.241 & 0.170 & 0.299 & -0.203 & -0.172 & -0.271 & 0.04 & 73.8 & 15.3 \\
274.0 & \verb|SXS:BBH:0307|(*) & 1.228 & - & - & 0.320 &-& - & -0.580 & -0.08 & 69.3 & 17.8 \\
273.9 & \verb|SXS:BBH:0312|(*) & 1.203 &-& - & 0.390 &-&-& -0.480 & -0.00 & 71.3 & 15.7 \\
273.6 & \verb|SXS:BBH:0308[Lev3]|(*) & 1.228 & 0.072 & 0.072 & 0.325 & 0.201 & 0.285 & -0.571 & -0.08 & 69.4 & 17.9 \\
273.5 & \verb|D21.5_q1_a0.2_0.8_th104.4775_n100| & 1.001 &-&-& 0.200 & 0.775 &-& -0.200 & -0.00 & 74.3 & 9.7 \\
273.5 & \verb|SXS:BBH:0003| & 1.0 & 0.497 & 0.053 & - &-&-&-& -0.00 & 73.8 & 11.4 \\
273.4 & \verb|SXS:BBH:0103| & 1.501 & 0.496 & 0.058 &-&-&-&-& 0.00 & 73.2 & 10.8 \\
273.4 & \verb|SXS:BBH:0310|(*) & 1.221 & - &-&-& - & - &-& 0.00 & 74.3 & 15.0 \\
273.1 & \verb|SXS:BBH:0015| & 1.501 & 0.487 & 0.110 & 0.001 &-&-&-& 0.00 & 74.5 & 10.5 \\
273.0 & \verb|SXS:BBH:0004| & 1.0 & - & - & -0.500 & - & - &-& -0.25 & 66.4 & 11.2 \\
273.0 & \verb|SXS:BBH:0024| & 1.501 & 0.496 & 0.051 & -0.001 & -0.077 & 0.489 & 0.002 & 0.00 & 73.2 & 12.5 \\
272.9 & \verb|SXS:BBH:0198| & 1.202 & - & - & - & - & - &-& 0.00 & 71.2 & 13.3 \\
272.8 & \verb|GT:BBH:900|(*) & 1.2 & 0.400 &-&-& 0.400 &-&-& 0.00 & 71.0 & 22.8 \\
272.8 & \verb|SXS:BBH:0123| & 1.1 & 0.267 & 0.020 & -0.415 & 0.038 & -0.054 & 0.126 & -0.16 & 69.9 & 15.9 \\
272.8 & \verb|SXS:BBH:0021| & 1.5 & 0.496 & 0.053 & - & - & 0.001 & -0.499 & -0.20 & 69.8 & 12.8 \\
272.7 & \verb|SXS:BBH:0147| & 1.0 & 0.404 & 0.294 & -0.001 & -0.404 & -0.294 & -0.001 & -0.00 & 71.5 & 24.4 \\
272.6 & \verb|SXS:BBH:0127| & 1.34 & 0.010 & -0.077 & -0.017 & -0.061 & -0.065 & -0.179 & -0.09 & 72.4 & 14.5 \\
272.6 & \verb|D11_q0.75_a0.6_0.6_PNr500_th1d_n100|(*) & 1.333 & 0.460 & 0.351 & 0.161 & 0.526 & 0.288 & -0.023 & 0.08 & 76.2 & 21.6 \\
272.5 & \verb|UD_D10.01_q1.00_a0.4_n100| & 1.0 &-&-& 0.400 &-&-& -0.400 & -0.00 & 74.6 & 26.2 \\
272.4 & \verb|D11_q0.75_a-0.5_0.5_n100|(*) & 1.333 &-&-& 0.500 &-&-& -0.500 & 0.07 & 74.8 & 22.1 \\
272.2 & \verb|SXS:BBH:0115| & 1.07 & 0.019 & 0.013 & -0.204 & 0.243 & -0.067 & 0.291 & 0.04 & 71.8 & 15.0 \\
272.2 & \verb|SXS:BBH:0308| & 1.228 & 0.094 & 0.056 & 0.322 & 0.266 & 0.213 & -0.576 & -0.08 & 72.9 & 17.0 \\
272.1 & \verb|SXS:BBH:0137| & 1.76 & -0.248 & -0.319 & -0.034 & -0.071 & 0.151 & -0.190 & -0.09 & 72.2 & 14.8 \\
272.0 & \verb|SXS:BBH:0125| & 1.27 & 0.012 & 0.045 & -0.058 & 0.389 & 0.241 & 0.070 & -0.00 & 71.6 & 13.9 \\
272.0 & \verb|D11_q0.75_a-0.5_0.0_n100|(*) & 1.333 &-&-& - &-&-& -0.500 & -0.21 & 68.7 & 24.1 \\
271.8 & \verb|SXS:BBH:0120| & 1.12 & 0.138 & -0.200 & -0.008 & -0.065 & -0.241 & -0.099 & -0.05 & 69.7 & 14.9 \\
271.8 & \verb|GT:BBH:448| & 1.0 &-&-&-&-&-&-& 0.00 & 75.6 & 21.6 \\
271.8 & \verb|GT:BBH:448| & 1.0 &-&-&-&-&-&-& 0.00 & 75.6 & 21.6 \\
271.8 & \verb|D12.00_q0.85_a0.0_0.0_n100|(*) & 1.176 &-&-& - &-&-& - & -0.00 & 71.3 & 21.1 \\
271.7 & \verb|D11_q0.75_a-0.8500_0.6375_n100| & 1.334 &-&-& 0.638 & 0.001 & 0.003 & -0.851 & -0.00 & 76.2 & 22.2 \\
271.6 & \verb|SXS:BBH:0138| & 1.7 & -0.044 & 0.425 & 0.042 & -0.012 & -0.008 & -0.111 & -0.01 & 72.5 & 14.4 \\
271.6 & \verb|D12_q1.00_a-0.25_0.25_n100|(*) & 1.0 &-&-& 0.250 &-&-& -0.250 & -0.00 & 70.9 & 21.2 \\
271.6 & \verb|D11_q0.75_a0.8_0.4_PNr500_th1d_n100|(*) & 1.334 & 0.074 & -0.374 & 0.123 & -0.420 & -0.531 & -0.427 & -0.11 & 70.3 & 24.0 \\
271.5 & \verb|SXS:BBH:0116| & 1.08 & -0.078 & 0.065 & 0.033 & 0.185 & 0.007 & 0.103 & 0.07 & 73.2 & 14.3 \\
271.3 & \verb|GT:BBH:370| & 1.15 &-&-&-&-&-&-& 0.00 & 75.5 & 25.3 \\
271.3 & \verb|GT:BBH:370| & 1.15 &-&-&-&-&-&-& 0.00 & 75.5 & 25.3 \\
271.2 & \verb|D10_q0.75_a-0.5_0.25_n100|(*) & 1.333 &-&-& 0.250 &-&-& -0.500 & -0.07 & 73.6 & 26.8 \\
271.2 & \verb|SXS:BBH:0010| & 1.501 & 0.248 & 0.028 & -0.433 & - &-&-& -0.26 & 65.8 & 14.1 \\
271.1 & \verb|GT:BBH:476| & 1.0 &-&-& -0.200 &-&-& -0.200 & -0.20 & 69.2 & 23.8 \\
271.1 & \verb|GT:BBH:476| & 1.0 &-&-& -0.200 &-&-& -0.200 & -0.20 & 69.2 & 23.8 \\
271.1 & \verb|SXS:BBH:0133| & 1.63 & 0.098 & 0.042 & -0.134 & -0.107 & -0.110 & -0.021 & -0.09 & 72.3 & 13.9 \\
271.1 & \verb|D12_q1.00_a-0.25_0.00_n100|(*) & 1.0 &-&-& - &-&-& -0.250 & -0.12 & 67.6 & 22.4 \\
\end{longtable}

\clearpage

\centerline{\large \textbf{Authors}}

\noindent
B.~P.~Abbott,$^{1}$  %
R.~Abbott,$^{1}$  %
T.~D.~Abbott,$^{2}$  %
M.~R.~Abernathy,$^{3}$  %
F.~Acernese,$^{4,5}$ %
K.~Ackley,$^{6}$  %
C.~Adams,$^{7}$  %
T.~Adams,$^{8}$ %
P.~Addesso,$^{9}$  %
R.~X.~Adhikari,$^{1}$
% ...

\medskip
\parindent 0pt
{$^{\dag}$Deceased, May 2015. }%
{$^{\ddag}$Deceased, March 2015. }%
\medskip

$^{1}$LIGO, California Institute of Technology, Pasadena, CA 91125, USA %
 
$^{2}$Louisiana State University, Baton Rouge, LA 70803, USA %
 
$^{3}$American University, Washington, D.C. 20016, USA %
 
$^{4}$Universit\`a di Salerno, Fisciano, I-84084 Salerno, Italy %
 
$^{5}$INFN, Sezione di Napoli, Complesso Universitario di Monte S.Angelo, I-80126 Napoli, Italy %
 
$^{6}$University of Florida, Gainesville, FL 32611, USA %
 
$^{7}$LIGO Livingston Observatory, Livingston, LA 70754, USA %
 
$^{8}$Laboratoire d'Annecy-le-Vieux de Physique des Particules (LAPP), Universit\'e Savoie Mont Blanc, CNRS/IN2P3, F-74941 Annecy-le-Vieux, France %
 
$^{9}$University of Sannio at Benevento, I-82100 Benevento, Italy and INFN, Sezione di Napoli, I-80100 Napoli, Italy %
 
$^{10}$Albert-Einstein-Institut, Max-Planck-Institut f\"ur Gravi\-ta\-tions\-physik, D-30167 Hannover, Germany %
 
$^{11}$Nikhef, Science Park, 1098 XG Amsterdam, The Netherlands %
 
$^{12}$LIGO, Massachusetts Institute of Technology, Cambridge, MA 02139, USA %
 
$^{13}$Instituto Nacional de Pesquisas Espaciais, 12227-010 S\~{a}o Jos\'{e} dos Campos, S\~{a}o Paulo, Brazil %
 
$^{14}$INFN, Gran Sasso Science Institute, I-67100 L'Aquila, Italy %
 
$^{15}$INFN, Sezione di Roma Tor Vergata, I-00133 Roma, Italy %
 
$^{16}$Inter-University Centre for Astronomy and Astrophysics, Pune 411007, India %
 
$^{17}$International Centre for Theoretical Sciences, Tata Institute of Fundamental Research, Bangalore 560012, India %
 
$^{18}$University of Wisconsin-Milwaukee, Milwaukee, WI 53201, USA %
 
$^{19}$Leibniz Universit\"at Hannover, D-30167 Hannover, Germany %
 
$^{20}$Universit\`a di Pisa, I-56127 Pisa, Italy %
 
$^{21}$INFN, Sezione di Pisa, I-56127 Pisa, Italy %
 
$^{22}$Australian National University, Canberra, Australian Capital Territory 0200, Australia %
 
$^{23}$The University of Mississippi, University, MS 38677, USA %
 
$^{24}$California State University Fullerton, Fullerton, CA 92831, USA %
 
$^{25}$LAL, Univ. Paris-Sud, CNRS/IN2P3, Universit\'e Paris-Saclay, Orsay, France %
 
$^{26}$Chennai Mathematical Institute, Chennai 603103, India %
 
$^{27}$Universit\`a di Roma Tor Vergata, I-00133 Roma, Italy %
 
$^{28}$University of Southampton, Southampton SO17 1BJ, United Kingdom %
 
$^{29}$Universit\"at Hamburg, D-22761 Hamburg, Germany %
 
$^{30}$INFN, Sezione di Roma, I-00185 Roma, Italy %
 
$^{31}$Albert-Einstein-Institut, Max-Planck-Institut f\"ur Gravitations\-physik, D-14476 Potsdam-Golm, Germany %
 
$^{32}$APC, AstroParticule et Cosmologie, Universit\'e Paris Diderot, CNRS/IN2P3, CEA/Irfu, Observatoire de Paris, Sorbonne Paris Cit\'e, F-75205 Paris Cedex 13, France %
 
$^{33}$Montana State University, Bozeman, MT 59717, USA %
 
$^{34}$Universit\`a di Perugia, I-06123 Perugia, Italy %
 
$^{35}$INFN, Sezione di Perugia, I-06123 Perugia, Italy %
 
$^{36}$European Gravitational Observatory (EGO), I-56021 Cascina, Pisa, Italy %
 
$^{37}$Syracuse University, Syracuse, NY 13244, USA %
 
$^{38}$SUPA, University of Glasgow, Glasgow G12 8QQ, United Kingdom %
 
$^{39}$LIGO Hanford Observatory, Richland, WA 99352, USA %
 
$^{40}$Wigner RCP, RMKI, H-1121 Budapest, Konkoly Thege Mikl\'os \'ut 29-33, Hungary %
 
$^{41}$Columbia University, New York, NY 10027, USA %
 
$^{42}$Stanford University, Stanford, CA 94305, USA %
 
$^{43}$Universit\`a di Padova, Dipartimento di Fisica e Astronomia, I-35131 Padova, Italy %
 
$^{44}$INFN, Sezione di Padova, I-35131 Padova, Italy %
 
$^{45}$CAMK-PAN, 00-716 Warsaw, Poland %
 
$^{46}$University of Birmingham, Birmingham B15 2TT, United Kingdom %
 
$^{47}$Universit\`a degli Studi di Genova, I-16146 Genova, Italy %
 
$^{48}$INFN, Sezione di Genova, I-16146 Genova, Italy %
 
$^{49}$RRCAT, Indore MP 452013, India %
 
$^{50}$Faculty of Physics, Lomonosov Moscow State University, Moscow 119991, Russia %
 
$^{51}$SUPA, University of the West of Scotland, Paisley PA1 2BE, United Kingdom %
 
$^{52}$University of Western Australia, Crawley, Western Australia 6009, Australia %
 
$^{53}$Department of Astrophysics/IMAPP, Radboud University Nijmegen, P.O. Box 9010, 6500 GL Nijmegen, The Netherlands %
 
$^{54}$Artemis, Universit\'e C\^ote d'Azur, CNRS, Observatoire C\^ote d'Azur, CS 34229, Nice cedex 4, France %
 
$^{55}$Institut de Physique de Rennes, CNRS, Universit\'e de Rennes 1, F-35042 Rennes, France %
 
$^{56}$Washington State University, Pullman, WA 99164, USA %
 
$^{57}$Universit\`a degli Studi di Urbino ``Carlo Bo,'' I-61029 Urbino, Italy %
 
$^{58}$INFN, Sezione di Firenze, I-50019 Sesto Fiorentino, Firenze, Italy %
 
$^{59}$University of Oregon, Eugene, OR 97403, USA %
 
$^{60}$Laboratoire Kastler Brossel, UPMC-Sorbonne Universit\'es, CNRS, ENS-PSL Research University, Coll\`ege de France, F-75005 Paris, France %
 
$^{61}$Carleton College, Northfield, MN 55057, USA %
 
$^{62}$Astronomical Observatory Warsaw University, 00-478 Warsaw, Poland %
 
$^{63}$VU University Amsterdam, 1081 HV Amsterdam, The Netherlands %
 
$^{64}$University of Maryland, College Park, MD 20742, USA %
 
$^{65}$Center for Relativistic Astrophysics and School of Physics, Georgia Institute of Technology, Atlanta, GA 30332, USA %
 
$^{66}$Laboratoire des Mat\'eriaux Avanc\'es (LMA), CNRS/IN2P3, F-69622 Villeurbanne, France %
 
$^{67}$Universit\'e Claude Bernard Lyon 1, F-69622 Villeurbanne, France %
 
$^{68}$Universit\`a di Napoli ``Federico II,'' Complesso Universitario di Monte S.Angelo, I-80126 Napoli, Italy %
 
$^{69}$NASA/Goddard Space Flight Center, Greenbelt, MD 20771, USA %
 
$^{70}$RESCEU, University of Tokyo, Tokyo, 113-0033, Japan. %
 
$^{71}$Tsinghua University, Beijing 100084, China %
 
$^{72}$Texas Tech University, Lubbock, TX 79409, USA %
 
$^{73}$National Tsing Hua University, Hsinchu City, 30013 Taiwan, Republic of China %
 
$^{74}$Charles Sturt University, Wagga Wagga, New South Wales 2678, Australia %
 
$^{75}$West Virginia University, Morgantown, WV 26506, USA %
 
$^{76}$University of Chicago, Chicago, IL 60637, USA %
 
$^{77}$Caltech CaRT, Pasadena, CA 91125, USA %
 
$^{78}$Korea Institute of Science and Technology Information, Daejeon 305-806, Korea %
 
$^{79}$Universit\`a di Roma ``La Sapienza,'' I-00185 Roma, Italy %
 
$^{80}$University of Brussels, Brussels 1050, Belgium %
 
$^{81}$Sonoma State University, Rohnert Park, CA 94928, USA %
 
$^{82}$Center for Interdisciplinary Exploration \& Research in Astrophysics (CIERA), Northwestern University, Evanston, IL 60208, USA %
 
$^{83}$University of Minnesota, Minneapolis, MN 55455, USA %
 
$^{84}$The University of Melbourne, Parkville, Victoria 3010, Australia %
 
$^{85}$Institute for Plasma Research, Bhat, Gandhinagar 382428, India %
 
$^{86}$The University of Sheffield, Sheffield S10 2TN, United Kingdom %
 
$^{87}$The University of Texas Rio Grande Valley, Brownsville, TX 78520, USA %
 
$^{88}$Universit\`a di Trento, Dipartimento di Fisica, I-38123 Povo, Trento, Italy %
 
$^{89}$INFN, Trento Institute for Fundamental Physics and Applications, I-38123 Povo, Trento, Italy %
 
$^{90}$The Pennsylvania State University, University Park, PA 16802, USA %
 
$^{91}$Cardiff University, Cardiff CF24 3AA, United Kingdom %
 
$^{92}$Montclair State University, Montclair, NJ 07043, USA %
 
$^{93}$MTA E\"otv\"os University, ``Lendulet'' Astrophysics Research Group, Budapest 1117, Hungary %
 
$^{94}$National Astronomical Observatory of Japan, 2-21-1 Osawa, Mitaka, Tokyo 181-8588, Japan %
 
$^{95}$School of Mathematics, University of Edinburgh, Edinburgh EH9 3FD, United Kingdom %
 
$^{96}$Indian Institute of Technology, Gandhinagar Ahmedabad Gujarat 382424, India %
 
$^{97}$University of Szeged, D\'om t\'er 9, Szeged 6720, Hungary %
 
$^{98}$Embry-Riddle Aeronautical University, Prescott, AZ 86301, USA %
 
$^{99}$Tata Institute of Fundamental Research, Mumbai 400005, India %
 
$^{100}$INAF, Osservatorio Astronomico di Capodimonte, I-80131, Napoli, Italy %
 
$^{101}$University of Michigan, Ann Arbor, MI 48109, USA %
 
$^{102}$Rochester Institute of Technology, Rochester, NY 14623, USA %
 
$^{103}$NCSA, University of Illinois at Urbana-Champaign, Urbana, Illinois 61801, USA %
 
$^{104}$University of Bia{\l }ystok, 15-424 Bia{\l }ystok, Poland %
 
$^{105}$SUPA, University of Strathclyde, Glasgow G1 1XQ, United Kingdom %
 
$^{106}$Universitat de les Illes Balears, IAC3---IEEC, E-07122 Palma de Mallorca, Spain %
 
$^{107}$IISER-TVM, CET Campus, Trivandrum Kerala 695016, India %
 
$^{108}$Canadian Institute for Theoretical Astrophysics, University of Toronto, Toronto, Ontario M5S 3H8, Canada %
 
$^{109}$Institute of Applied Physics, Nizhny Novgorod, 603950, Russia %
 
$^{110}$Pusan National University, Busan 609-735, Korea %
 
$^{111}$Hanyang University, Seoul 133-791, Korea %
 
$^{112}$University of Adelaide, Adelaide, South Australia 5005, Australia %
 
$^{113}$NCBJ, 05-400 \'Swierk-Otwock, Poland %
 
$^{114}$IM-PAN, 00-956 Warsaw, Poland %
 
$^{115}$Monash University, Victoria 3800, Australia %
 
$^{116}$Seoul National University, Seoul 151-742, Korea %
 
$^{117}$The Chinese University of Hong Kong, Shatin, NT, Hong Kong SAR, China %
 
$^{118}$University of Alabama in Huntsville, Huntsville, AL 35899, USA %
 
$^{119}$University of Massachusetts-Amherst, Amherst, MA 01003, USA %
 
$^{120}$ESPCI, CNRS, F-75005 Paris, France %
 
$^{121}$Universit\`a di Camerino, Dipartimento di Fisica, I-62032 Camerino, Italy %
 
$^{122}$Southern University and A\&M College, Baton Rouge, LA 70813, USA %
 
$^{123}$College of William and Mary, Williamsburg, VA 23187, USA %
 
$^{124}$Instituto de F\'\i sica Te\'orica, University Estadual Paulista/ICTP South American Institute for Fundamental Research, S\~ao Paulo SP 01140-070, Brazil %
 
$^{125}$University of Cambridge, Cambridge CB2 1TN, United Kingdom %
 
$^{126}$IISER-Kolkata, Mohanpur, West Bengal 741252, India %
 
$^{127}$Rutherford Appleton Laboratory, HSIC, Chilton, Didcot, Oxon OX11 0QX, United Kingdom %
 
$^{128}$Whitman College, 345 Boyer Avenue, Walla Walla, WA 99362 USA %
 
$^{129}$National Institute for Mathematical Sciences, Daejeon 305-390, Korea %
 
$^{130}$Universit\'e de Lyon, F-69361 Lyon, France %
 
$^{131}$Hobart and William Smith Colleges, Geneva, NY 14456, USA %
 
$^{132}$Janusz Gil Institute of Astronomy, University of Zielona G\'ora, 65-265 Zielona G\'ora, Poland %
 
$^{133}$King's College London, University of London, London WC2R 2LS, United Kingdom %
 
$^{134}$Andrews University, Berrien Springs, MI 49104, USA %
 
$^{135}$Universit\`a di Siena, I-53100 Siena, Italy %
 
$^{136}$Trinity University, San Antonio, TX 78212, USA %
 
$^{137}$University of Washington, Seattle, WA 98195, USA %
 
$^{138}$Kenyon College, Gambier, OH 43022, USA %
 
$^{139}$Abilene Christian University, Abilene, TX 79699, USA %
 
$^{140}$Cornell Center for Astrophysics and Planetary Science, Cornell University, Ithaca, NY 14853, USA %

\end{widetext}

\end{document}